\documentclass[usenatbib]{mnras}

\usepackage{graphicx,rotating}
\usepackage{amssymb,amsmath,pdflscape}
\usepackage{natbib}
\usepackage{subeqnarray}
\usepackage{mathtools}
\usepackage{cases}
\usepackage[utf8x]{inputenc}
\usepackage{auto-pst-pdf}
\usepackage{hyperref}
\usepackage{cleveref}

\newcommand{\pc}{p_{\rm c}}
\newcommand{\pint}{p_1}
\newcommand{\pext}{p_2}
\newcommand{\const}{{\rm const.}}
\newcommand{\psiint}{\Psi_{\rm int.}}
\newcommand{\psiext}{\Psi_{\rm ext.}}

\newcommand{\bareone}{\bar{\epsilon}_1}
\newcommand{\baretwo}{\bar{\epsilon}_2}

\newcommand{\ellenv}{\epsilon_2}
\newcommand{\ellcor}{\epsilon_1}

\newcommand{\alphac}{\alpha_{\rm C}}

\newcommand{\rhoenv}{\rho_2}

\begin{document}

\title[]{Nested spheroidal figures of equilibrium \\ I. Approximate solutions for rigid rotations}

\author[J.-M. Hur\'e]
       {J.-M. Hur\'e$^{1,2}$\thanks{E-mail:jean-marc.hure@u-bordeaux.fr}\\
$^{1}$Univ. Bordeaux, LAB, UMR 5804, F-33615, Pessac, France\\
$^{2}$CNRS, LAB, UMR 5804, F-33615, Pessac, France}

\date{Received ??? / Accepted ???}
 
\pagerange{\pageref{firstpage}--\pageref{lastpage}} \pubyear{???}

\maketitle

\label{firstpage}

\begin{abstract}
We discuss the equilibrium conditions for a body made of two homogeneous components separated by oblate spheroidal surfaces and in relative motion. While exact solutions are not permitted for rigid rotation (unless a specific ambient pressure), approximations can be obtained for configurations involving a small confocal parameter. The problem then admits two families of solutions, depending on the pressure along the common interface (constant or quadratic with the cylindrical radius). We give in both cases the pressure and the rotation rates as a function of the fractional radius, ellipticities and mass-density jump. Various degrees of flattening are allowed but there are severe limitations for global rotation, as already known from classical theory (e.g. impossibility of confocal and coelliptical solutions, gradient of ellipticity outward). States of relative rotation are much less constrained, but these require a mass-density jump. This analytical approach compares successfully with the numerical solutions obtained from the Self-Consistent-Field method. Practical formula are derived in the limit of small ellipticities appropriate for slowly-rotating star/planet interiors.
\end{abstract}

\begin{keywords}
Gravitation | stars: interiors | stars: rotation | planets and satellites: interiors | Methods: analytical
\end{keywords}

\section{Introduction}

According to the theory of figures \citep[][]{chandra69}, a homogeneous body bounded by a spheroidal surface $E(a,b)$ with semi-minor axis $b$ and semi-major axis $a \ge b = \bar{\epsilon} a$ is in self-gravitating equilibrium if the rotation rate $\Omega$ and the mass density $\rho$ are linked by
\begin{flalign}
\frac{\Omega^2}{2\pi G \rho} = {\cal M}(\epsilon),
\label{eq:omega2ref}
\end{flalign}
where $\epsilon=\sqrt{1-\bar{\epsilon}^2}$ is the ellipticity, $G$ is the constant of gravitation, and
\begin{flalign}
{\cal M}(\epsilon)=\left(3-2\epsilon^2\right)\frac{\bar{\epsilon}}{\epsilon^3}\arcsin(\epsilon)+3-\frac{3}{\epsilon^2} \ge 0.
\label{eq:mlfunction}
\end{flalign}
This result is due to Maclaurin.
In which conditions a body made of two rotating components separated by spheroidal surfaces can be a figure of equilibrium, and what types of solutions are permitted? These questions have been examined in details more than one century ago by Hamy, Poincar\'e and others in the specific context of the Earth and planets, already in the multilayer case \citep{alma990001099760306161}; see also \cite{pohanka11} and \cite{ragazzo2018theory}. As most systems in the Universe are significatly flattened by rotation and exhibit a certain internal stratification, the problem is obviously of wider interest. The interior of stars is concerned, with, for instance, the discovery of limits in the mass and size of nuclear cores \citep{sc42,maeder71,ru88,roze01,kiu10,ka16}. Galactic halos hostings disks represent another class of composite systems, although mainly non-collisional, the equilibrium and stability of which have been studied from theory of ellipsoidal figures \citep{ak74,ak75,du78,rl84,sm86,cs90,mcm90}.

The establishment of the conditions for the existence of nested, spheroidal figures is a tricky problem, even under the assumptions of incompressibility and rigid rotation. Actually, with only two components, this is already a four dimensional problem. \cite{poincare88} proved that only configurations involving confocal spheroidal surfaces are viable if all layers rotate synchroneously, while \cite{hamy90} pointed out that such states require a density inversion (the mass density increases from centre to surface); see also \cite{mmc83}. In this article, we present a concise survey of the $2$-layer problem and focus on oblate spheroidal, bounding surfaces (in the meridional plane, the layers are separated by perfect ellipses). The more complicated case of a multilayer system is treated in a forthcoming article (paper II). The two components are homogeneous and treated as collisional fluids in the sense that the two phases do not mix. In contrast with most previous investigations, we initially relax the assumptions of confocality, coellipticity and common rotation rate. The question of the rotation laws is central. According to Poincar\'e's and Hamy's theorems \citep{poincare88,moulton16}, global rotation is not permitted neither for coelliptical nor for confocal configurations. Nevertheless, the numerical approach  \citep{bh21} shows that, for global rotation, i) the ``cores'' are generally more spherical than the surrounding layers, and ii) the bounding surfaces are remarkably close to ellipses. This is observed not only in the incompressible case but for a wide range of polytropic indices. This article is therefore partly motivated by this apparent paradox. In a remarkable monograph, \cite{hamy89} has shown that rigid rotation is verified in a first approximation for small ellipticties. Here, we give an extended version of this analysis based on an expansion in the confocal parameter (rather than in the ellipticities). Besides, we wish to go beyond the ``simple'' case of global rotation and seek for solutions involving asynchroneously rotating layers.

There are different options to treat the problem: i) directly from the three components of forces \cite[e.g.][]{hamy89}, ii) from the modified Lane-Emden equation \citep{cai86}, iii) from the conservation of energy \citep{mmc83}, and iv) from the Virial equations, which is probably the most powerful approach \citep[e.g.][]{chandra69,maeder71,ak74,du78,bcs83,cs90}. In this paper, we follow the second path, and directly make use of the Bernoulli equation. We show that, to the first order in the confocal parameter (see below), the problem admits solutions compatible with the rigid rotation law. This is discussed in Sects. \ref{sec:equi} and \ref{sec:lambda}. We then examine the case of global rotation (type-C solutions; same rotation rate for both components) in Sect. \ref{sec:typec}. We consider the more general situation where the embedded spheroid and the surrounding layer are in relative motion (type-V solutions) in Sect. \ref{sec:typev}. A few practical formula valid in the limit of small ellipticities are derived in Sect. \ref{sec:srl}. Exact solutions corresponding to differentially rotating components are given in the concluding section. A basic (non-optimized) F90-code is appended.

\section{The equations of equilibrium}
\label{sec:equi}

\subsection{Hypothesis and notations}
\label{sec:hyp}

Inside a Maclaurin spheroid (see the Introduction), the pressure of matter $p$ varies according to the Bernoulli equation
\begin{flalign}
  \frac{p}{\rho} - \frac{1}{2}\Omega^2 R^2 + \psiint = \const,
  \label{eq:bernoulli}
\end{flalign}
where the constant can be evaluated at the center of coordinates, and $\psiint$ is the interior gravitational potential. In polar cylindrical coodinates $(R,Z)$, this function writes
\begin{flalign}
\frac{\psiint(R,Z)}{-\pi G \rho}= A_0(\epsilon)a^2-A_1(\epsilon)R^2 -A_3(\epsilon)Z^2
  \label{eq:psiint},
\end{flalign}
where
\begin{align}
  \begin{dcases}
    A_0(\epsilon)=2\frac{\bar{\epsilon}}{\epsilon}\arcsin \epsilon,\\
    A_1(\epsilon)=\frac{\bar{\epsilon}}{\epsilon^3}\left[\arcsin\epsilon-\epsilon\bar{\epsilon}\right],\\
    A_3(\epsilon)=-2\frac{\bar{\epsilon}}{\epsilon^3}\left[\arcsin\epsilon-\frac{\epsilon}{\bar{\epsilon}}\right],
    \label{eq:IA1A3}
  \end{dcases}
  \end{align}
and $\bar{\epsilon}=b/a$ is the dimensionless polar radius. Note that $A_1(\epsilon)-\bar{\epsilon}^2 A_3(\epsilon) \equiv {\cal M}(\epsilon)$; see (\ref{eq:mlfunction}). The knowledge of the external potential is essential to treat the problem of nested figures. It is given by \citep[e.g.][]{chandra69,binneytremaine87}
\begin{flalign}
  \nonumber
  \frac{\psiext(R,Z)}{-\pi G \rho }=  f &\left[ A_0(\epsilon')(a^2+\lambda) \right.\\
    & \qquad \left.-A_1(\epsilon')R^2 -A_3(\epsilon')Z^2 \right],
  \label{eq:psiext}
\end{flalign}
where the $A_i$'s are still given by (\ref{eq:IA1A3}),
\begin{subnumcases}{}
  \frac{R^2}{a^2+\lambda}+  \frac{Z^2}{b^2+\lambda} -1 =0 \label{eq:Asa},\\
 f=\frac{a^2b}{(a^2+\lambda)\sqrt{b^2+\lambda}}, \label{eq:Asb}\\
 {\epsilon'}^2 = 1 - \frac{b^2+\lambda}{a^2+\lambda}.  \label{eq:Asc}
\end{subnumcases}
In (\ref{eq:Asa}), $\lambda$ is clearly the root of a second-degree polynomial. In the present case, only the largest positive root is relevant. For a given pair $(a,b)$, this quantity varies with $R$ and $Z$; see Sect. \ref{sec:lambda}. It can be verified that (\ref{eq:psiint}) and (\ref{eq:psiext}) coincide at the boundary $E$, where $\lambda=0$.

\begin{figure}
\includegraphics[width=8.6cm,bb=0 0 575 549,clip==]{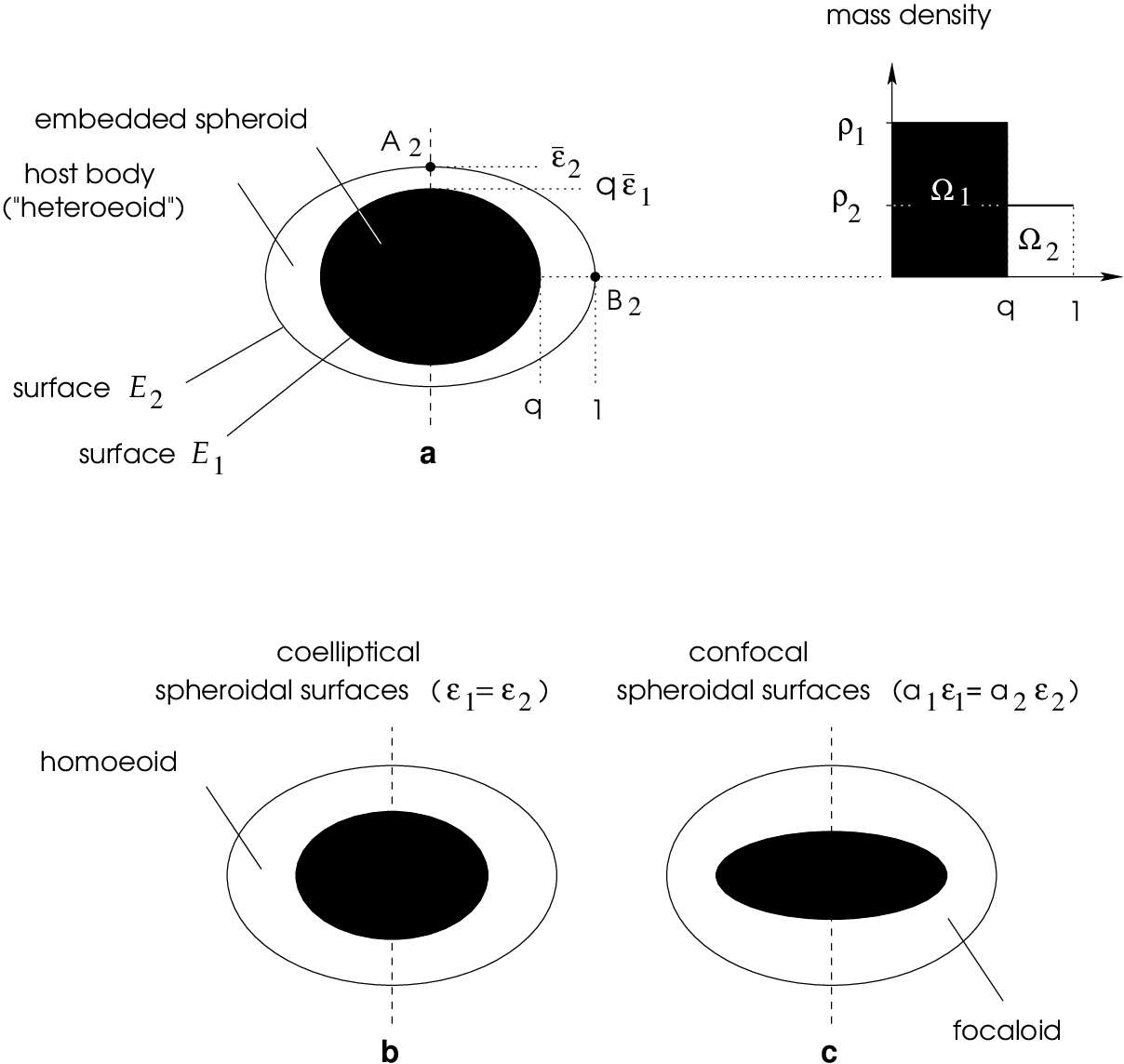}
\caption{Typical configurations for a composite system made of a homogeneous spheroid ({\it black}) surrounded by a hollow body ({\it white}), the ``host'' bounded by two spheroidal surfaces. The embedded spheroid and the host are treated as collisional fluids. Each component rotate around the $Z$-axis ({\it dashed line)} at its own rate. Two special cases are depicted : $E_1$ and $E_2$ are similar/homothetic (b),  $E_1$ and $E_2$ are confocal (c).}
\label{fig:configsVCbis.eps}
\end{figure}

    We consider that this spheroid (we use the subscript $1$ for associated quantities), is embedded inside a homogeneous body (subscript $2$), called the ``host'', which shares the same axis of revolution and same plane of symmetry (and subsequently, the same center). This is a hollow body, internally bounded by $E_1$ (i.e. the common interface) and externally bounded by a larger, spheroidal surface $E_2(a_2,b_2)$, with semi-minor axis $b_2$ and semi-major axis $a_2 \ge b_2= \baretwo a_2$. The spheroidal surfaces are not allowed to intersect. This two-component system is depicted in Fig. \ref{fig:configsVCbis.eps}a. Another important assumption is that there is no exchange or travel of matter between the two components. In a collisionaless system, as in galaxies, this permitted \citep[e.g.][]{ak75,cai86}.

According to the common convention \citep{kelvin1883treatise,perek50}, the host is called a ``focaloid'' when  $E_1$ and $E_2$ are confocal, which corresponds to $a_1\epsilon_1=a_2\epsilon_2$. It is called a ``homoeoid'' when $E_1$ and $E_2$ are similar or homothetic surfaces, which means $\epsilon_1=\epsilon_2$. We can introduce the word ``heteroeoid'' to refer to the general case where $E_1$ and $E_2$ are neither confocal nor similar (but still not intersecting). Although a selection of preferred configurations will be perfomed, we do not impose any constraint yet on the ellipticities $\epsilon_1$ and $\epsilon_2$, and each one can run over the full range $[0,1]$. If $q=a_1/a_2$ denotes the fractional size of the embedded spheroid, the condition of complete immersion writes
\begin{flalign}
  \begin{cases}
\baretwo- q \bareone \ge 0,\\
   q \le 1.
 \label{eq:immersion}
 \end{cases}
\end{flalign}

An important ingredient of the problem is the mass-density jump\footnote{The density contrast of the nested spheroid with respect to the host is $\alpha-1 \equiv \frac{\delta \rho}{\rho_2}$.} defined as
  \begin{equation}
    \alpha=\frac{\rho_1}{\rho_2}.
  \end{equation}
 It must be a positive constant, with a preference for
\begin{flalign}
  \alpha >1,
 \label{alphalargerthanone}
\end{flalign}
otherwise the host is more dense than the embedded body (density inversion). With these notations, the total volume of the system is $V=\frac{4}{3}\pi a_2^3 \baretwo$ and the total mass $M$ is given by
  \begin{flalign}
   M= \rho_2 V \left[1+(\alpha-1) q^3 \frac{\bareone}{\baretwo}\right],
\end{flalign}
  which leads to the mean mass-density $\bar{\rho}=M/V$ and to the fractional masses, namely
\begin{flalign}
  \nu_1=\frac{\alpha q^3\bareone}{\baretwo+(\alpha-1) q^3 \bareone}
\end{flalign}
 for the embedded spheroid and $\nu_2=1-\nu_1$ for the host.

 A great diversity of configurations is a priori conceivable, as depicted in Fig. \ref{fig:configsVCbis.eps}, since each component can, as long as the equilibrium conditions and (\ref{eq:immersion}) allow, occupy any state from a highly-flattened structure to a quasi-spherical one. 

\subsection{The equations of equilibrium}

Working with exact spheroidal surfaces offers a great mathematical simplification as it fixes the gravitational potential. The global equilibrium requires that (\ref{eq:bernoulli}) holds for each component. For the embedded spheroid, the relevant gravitational potential $\Psi_1$ is found from superposition by considering (\ref{eq:psiint}) with appropriate settings for the mass densities and for the $a$'s and $b$'s involved, namely
\begin{flalign}
\frac{\Psi_1(R,Z)}{-\pi G \rho_2}= A'_0-A'_1R^2 -A'_3Z^2
  \label{eq:psiint1},
\end{flalign}
where
\begin{flalign}
  \begin{cases}
  A'_0 = A_0(\epsilon_2)a_2^2 + (\alpha-1) A_0(\epsilon_1)a_1^2,\\
  A'_i = A_i(\epsilon_2) + (\alpha-1) A_i(\epsilon_1),  \qquad i \in \{1,3\}.
  \end{cases}
  \label{eq:aprim}
\end{flalign}
In these conditions, the equilibrium of the embedded spheroid, assumed in rigid rotation, is dictated by (\ref{eq:bernoulli}) and (\ref{eq:psiint1}), namely
\begin{flalign}
  \frac{\pint}{\rho_1} - \frac{1}{2}\Omega^2_1 R^2 + \pi G \rho_2 \left( A'_1 R^2 + A_3'Z^2 \right) =  \frac{\pc}{\rho_1},
  \label{eq:bernoulli1}
\end{flalign}
where $\pc \equiv p_1(0,0)$ is the central pressure. Inside the host, the gravitational potential $\Psi_2$ combines an interior form and an exterior form. From (\ref{eq:psiint}) and (\ref{eq:psiext}), we have
\begin{flalign}
\frac{\Psi_2(R,Z)}{-\pi G \rho_2}= A''_0-A''_1R^2 -A''_3Z^2,
  \label{eq:psiint2}
\end{flalign}
where
\begin{flalign}
  \begin{cases}
  A''_0 = A_0(\epsilon_2)a_2^2+ (\alpha-1) f_1 A_0\left(\epsilon'_1\right)(a_1^2+\lambda),&\\
  A''_i = A_i(\epsilon_2)+ (\alpha-1) f_1A_i(\epsilon'_1), \qquad i \in \{1,3\},&
  \end{cases}
  \label{eq:adoubleprim}
\end{flalign}
and $f_1$ and $\epsilon_1'$ are still given by (\ref{eq:Asb}) and (\ref{eq:Asc}) respectively, but for $(a,b)=(a_1,b_1)$. The $A_i''$'s depend clearly on three parameters, namely $\epsilon_1$, $\epsilon_2$ and $\alpha$, but, in contrast, with the $A_i'$'s, these quantities also depend on $\lambda$, like $f_1$ and $\epsilon'_1$, and subsequently on $R$ and $Z$. Again, assuming rigid rotation (at a rate $\Omega_2$), the Bernoulli equation for the host writes
\begin{flalign}
  \nonumber
  &\frac{\pext}{\rho_2} - \frac{1}{2}\Omega_2^2 R^2\\
  &\qquad \qquad -\pi G \rho_2\left( A''_0 - A''_1 R^2 -A_3''Z^2 \right) =  \const',
       \label{eq:bernoulli2}
\end{flalign}
where the constant can be derived at the surface. For instance at point A$_2(0,b_2)$ (see Fig. \ref{fig:configsVCbis.eps}), provided the ambient medium brings no contribution to the pressure, we have 
\begin{flalign}
\const' = -\pi G \rho_2 \left. \left(A''_0 -A''_3 b_2^2 \right)\right|_{{\rm A}_2},
\end{flalign}
where the values of $f_1$, $\epsilon_1'$ and $\lambda$ required at point A$_2$ are easily found (see below).

Another decisive equation is the requirement of pressure balance at the connection between the two components, namely
\begin{flalign}
  \left.\left(p_2-p_1\right)\right|_{E_1}=0.
  \label{eq:pbalance}
\end{flalign}
The continuity of the pressure and the continuity of the gravitational potential at $E_1$ means that the mass densities $\rho_1$ and $\rho_2$ on both sides of the interface must be coherent with the Bernoulli equations (\ref{eq:bernoulli1}) and (\ref{eq:bernoulli2}), which implies a jump, i.e. $\alpha \ne 1$. Eventually, a discontinuity in the rotation rates can occur in addition. From this point of view, the fact that $\Omega_2$ and $\Omega_1$ can differ {\it at a given radius} does not seem in contradiction with the Poincar\'e-Wavre theorem. Barotropic systems indeed require that the rotation rate must be constant on cylinders \citep{tassoul78,amendt1989}. Here, howewer, we have two components in contact and the change in the rotation law at the interface is associated with a coherent change in the mass density \citep[e.g.][]{mmc83,cai86,kiu10}.

\subsection{Note on the pressure on the polar axis}

A nested equilibrium is in principle found by solving (\ref{eq:bernoulli1}), (\ref{eq:bernoulli2}) and (\ref{eq:pbalance}), with the conditions (\ref{eq:immersion}) and (\ref{alphalargerthanone}). As the centrifugal forces vanish on the rotation axis $R=0$ (this should be true for rotation laws other than rigid), the pressure $p^*(E_1) \equiv p_1(0,b_1) \equiv p_2(0,b_1)$ at point A$_1$ of the polar axis and the central pressure $\pc \equiv p_1(0,0)$ can already be calculated. From (\ref{eq:bernoulli1}), (\ref{eq:bernoulli2}) and (\ref{eq:pbalance}), we actually find
\begin{flalign}
\pc =  p^*(E_1) + \pi G \rho_2^2 \alpha A_3' b_1^2,
  \label{eq:pc}
\end{flalign}
and
\begin{flalign}
  \frac{p^*(E_1)}{\pi G \rho_2^2}=A'_0 -A_3' b_1^2 - \left. \left(A''_0- A_3'' b_2^2\right)\right|_{{\rm A}_2}.
  \label{eq:pinterface}
\end{flalign}
These two expressions are exact \citep{rl84}.

\section{The $\lambda$-parameter out of confocality}
\label{sec:lambda}

\subsection{Conditions for approximate rigid rotations}

As outlined above, the $\lambda$-parameter is a key-quantity of the problem. While $E_1$ is characterized by $\lambda=0$, we see from (\ref{eq:Asa}) that $E_2$ generally involves a continuum of values for $\lambda$ ranging from $b_2^2-b_1^2 \equiv \lambda_{{\rm A}_2}$ (on the polar axis; see point A$_2$ in Fig. \ref{fig:configsVCbis.eps}a) to $a_2^2-a_1^2 \equiv \lambda_{{\rm B_2}}$ (at the equator, i.e. point B$_2$). If we define
\begin{flalign}
  x a_2^2=a_1^2+\lambda,
 \label{eq:x}
\end{flalign}
we have at the two end-points
\begin{flalign}
  \begin{cases}
    x_{{\rm A}_2}=1+c,\\
    x_{{\rm B}_2}=1,
 \label{eq:xab}
 \end{cases}
\end{flalign}
where
\begin{flalign}
  c=q^2 \epsilon_1^2 - \epsilon_2^2,
  \label{eq:confocalc}
\end{flalign}
is the ``confocal'' parameter. We then have 
\begin{flalign}
  \begin{dcases}
   \left. f_1\right|_{{\rm A}_2}=\frac{q^3\bareone}{(1+c)\sqrt{1+c-q^2\epsilon_1^2}},\\
    \left.{\epsilon'}\right|_{{\rm A}_2}=\frac{q\epsilon_1}{\sqrt{1+c}},\\
    \left. f_1\right|_{{\rm B}_2}=\frac{q^3\bareone}{\sqrt{1-q^2\epsilon_1^2}},\\
    \left.{\epsilon'}\right|_{{\rm B}_2}=q\epsilon_1.
 \label{eq:feprimab}
 \end{dcases}
\end{flalign}
Except if $E_2$ is confocal with $E_1$ (in which case $c=0$), we see from (\ref{eq:Asa}) and (\ref{eq:x}) that $x$ varies along the surface according to
\begin{flalign}
  x^2-(1+c+\epsilon_2^2 \varpi^2)x+(c+\epsilon_2^2) \varpi^2=0,
  \label{eq:p2x}
\end{flalign}
where
\begin{flalign}
  \varpi=\frac{R}{a_2} \in [0,1].
  \label{eq:varpi}
\end{flalign}
The relevant value for $x$ is the largest, positive root of this equation. It is given by
\begin{flalign}
  2x=1+c+\epsilon_2^2 \varpi^2+\sqrt{(1+c+\epsilon_2^2 \varpi^2)^2-4 q^2 \epsilon_1^2 \varpi^2},
  \label{eq:xexact}
\end{flalign}
which is clearly not constant with the cylindrical radius. It means that, at the surface of the system, the $A_i''$'s depend on $R$ and $Z$, with the consequence that (\ref{eq:bernoulli2}) cannot be satisfied for rigid rotation. But there are two exceptions: i) $E_1$ and $E_2$ are confocal, or ii) the ambient medium exerts a specific, non-constant pressure $p_a=p_2|_{E_2} >0$ along $E_2$. The first property is known for long \citep{poincare88}, but it implies $\alpha < 1$ \citep{hamy90,mmc83}; see below. Regarding the second one, $p_a$ must partially or totally neutralize the terms in $R^2$ coming from $\Psi_2$. Due to the continuity of the gravitational potential, the pressure inside the host can remain quadratic with the radius along any intermediate spheroidal surface located from $E_2$ down to $E_1$. Then, according to (\ref{eq:bernoulli1}) and (\ref{eq:pbalance}), the embedded spheroid can be in rigid rotation too, and the pressure inside can be either quadratic with the cylindrical radius or a constant. It follows that {\it a sufficient condition for the two homogeneous components of a heterogeneous systems separated by spheroidal surfaces be in rigid rotation is the existence of an ambient pressure.} Note that the origin of this ambient pressure is preferably a photon field, otherwise there would be an extra contribution to gravity that would affect $\Psi$.

\subsection{Orders zero and one in the confocal parameter}

It follows from the above discussion that, in the absence of any ambient pressure, {\it any solution obtained with rigid rotations must be regarded as an approximation}, as considered in \cite{hamy89}. Since the Bernoulli equation is already quadratic in $R$ and $Z$ (and $Z$ easily convertible in a function of $R$ on any ellipse), it is interesting to expand $x$ as a series of $\varpi^2$. In this purpose, the square root in (\ref{eq:xexact}) is rewritten as
\begin{flalign}
(1+c)  \sqrt{1+\frac{\varpi^2}{(1+c)^2}\left[2(1+c)\epsilon_2^2 - 4q^2 \epsilon_1^2 + \epsilon_2^4 \varpi^2 \right]},
\end{flalign}
and subsequently expanded. Because $\varpi \in [0,1]$, we have $2(1+c)\epsilon_2^2 - 4q^2 \epsilon_1^2 + \epsilon_2^4 \varpi^2 < (\epsilon_2^2-2) (\epsilon_2^2+2c)$. It follows that the term supporting $\varpi^2$ in the above expression is, in absolute, less than unity provided $|c| \ll 1$. This excludes infinitely flat configurations (i.e. disks). In these conditions, the solution $x$ can be put into the form
\begin{flalign}
  \nonumber
  x = 1 &+ c\left[1- \varpi^2 \left(1 - \frac{q^2\epsilon_1^2}{1+c} \right) \right.\\
    & \qquad \qquad \left.- \varpi^4 \frac{q^2\epsilon_1^2}{(1+c)^2}\left(1 - \frac{q^2\epsilon_1^2}{1+c} \right)  + \dots  \right],
  \label{eq:xexpansion}
\end{flalign}
where the factorization by $c$ is required as $x=1$ for $c=0$ for any $\varpi$. So, we have $x \approx 1$ at order zero in $|c| \ll 1$. As first-order, we retain the next terms compatible with (\ref{eq:xab}), namely
\begin{flalign}
  x \approx 1 + c(1- \varpi^2),
  \label{eq:xexpansionbis}
\end{flalign}
which implies i) $q^2\epsilon_1^2 \ll 1+c$ (i.e. $E_2$ is close to a sphere) or ii) $q^2 \ll 1$ (the embedded spheroid has small size) or iii) $\epsilon_1^2 \ll 1$ (the embedded spheroid is close to a sphere). Thus, these latter conditions do not necessarily imply that the two spheroidal surfaces $E_1$ and $E_2$ are simultaneously close to spheres. This is in contrast with respect to the classical appraoch \citep[e.g.][]{hamy89,cr63}.

We see that (\ref{eq:xexpansionbis}) is simple and attractive as it yields the correct values at the two end-points A$_2$ and B$_2$ of $E_2$; see (\ref{eq:xab}). Besides, (\ref{eq:xexpansion}) means that any regular function of $\lambda$ or of $x$ is an infinite series of $\varpi^2$. The $A_i''$'s which are needed in (\ref{eq:bernoulli2}) enter into this category. While we will make no use of such an information, it is interesting to see how these quantities depend on $\varpi$ along $E_2$ and on the confocal parameter $c$. If we Taylor-expand $A_i$ around the value at one of the end-points, for instance at point B$_2$ on the equator, we get
\begin{flalign}
A''_i = \left. A''_i\right|_{{\rm B}_2}+(x-x_{{\rm B}_2}) \left. \frac{\partial A''_i}{\partial x}\right|_{{\rm B}_2}+ \dots
\end{flalign}
Unfortunately, this formula, which has to be truncated in practice, does not lead to the required value on the polar axis (i.e. at point A$_2$ in Fig. \ref{fig:configsVCbis.eps}), but we already see that $x-x_{{\rm B}_2} \sim \varpi^2$. With a finite difference, we have
\begin{flalign}
A''_i \approx \left. A''_i\right|_{{\rm B}_2} + \frac{x-x_{{\rm B}_2}}{x_{{\rm A}_2}-x_{{\rm B}_2}}\left(\left. A''_i\right|_{{\rm A}_2}-\left. A''_i\right|_{{\rm B}_2}\right),
\label{eq:asec_approx}
\end{flalign}

\begin{flalign}
 \frac{x-x_{{\rm B}_2}}{x_{{\rm A}_2}-x_{{\rm B}_2}} = 1- \varpi^2.
 \end{flalign}
from (\ref{eq:x}) and (\ref{eq:xexpansion}). We therefore see that, at the lowest order in $c$, all the $A''_i$'s in (\ref{eq:bernoulli2}) remain unsensitive to the radius, and we have $x \approx x_{{\rm B}_2}=1$. To the first order in $c$, all the $A''_i$'s in (\ref{eq:bernoulli2}) bring a quadratic contribution in $\varpi$, which, depending on $\epsilon_1$, $\epsilon_2$ and $q$, reinforces or decreases the quadratic contribution explicitely present in (\ref{eq:psiint2}). Accordingly, $\Psi$ contains not only terms in $\varpi^2$ but also terms in $\varpi^4$ as well (see Sect. \ref{sec:conclusion}). The limit of rigid rotation is therefore reached. Whatever the variation of the $A''_i$'s along $E_2$, the values at the two-end points A$_2$ and B$_2$ are perfectly accessible from (\ref{eq:x}) and (\ref{eq:feprimab}), and we will use these values to go beyond order zero. Note that, if $c$ is indeed close to $0$, the derivatives $\partial A_i/\partial x$, and subsequently the term inside the parenthesis in the right-hand-side of (\ref{eq:asec_approx}), are expected to be small-amplitude corrections.

\section{Type-C solutions : the interface is a surface of constant pressure}
\label{sec:typec}

\subsection{Rotation rate and mass-density jump. Example}

We have now to determine the rotation rate for the two components and the conditions that these are real and positive values. We first focus on order zero (i.e. $x \approx 1$). The first family of solution is obtained by assuming that the interface $E_1$ is a surface of constant pressure \citep[e.g.][]{lyttleton1953stability}. On this surface, we have
\begin{flalign}
  Z^2=b_1^2\left(1-\frac{R^2}{a_1^2}\right).
\end{flalign}
As the $A_i'$ do not depend on $\lambda$, the rotation rate of the embedded spheroid is directly deduced from (\ref{eq:bernoulli1}), namely 
\begin{flalign}
  \frac{\Omega^2_1}{2\pi G \rho_2} = A'_1 - \bareone^2 A'_3,
  \label{eq:omega1}
\end{flalign}
which clearly differs from (\ref{eq:omega2ref}). This result is already known \citep[e.g.][]{ak74}. If, for convenience, we express the $\Omega^2$'s in units of $2\pi G \rho_2$ by setting
  \begin{flalign}
    \tilde{\Omega}_i \sqrt{2\pi G \rho_2}= \Omega_i,
    \label{eq:tildeomega}
\end{flalign}

\noindent and define
\begin{flalign}
  {\cal P}(\epsilon,\epsilon') = \frac{A_3(\epsilon')(1-\epsilon^2)-A_1(\epsilon')}{{\cal M}(\epsilon)},
 \label{eq:pfunction}
\end{flalign}
then (\ref{eq:omega1}) reads
\begin{flalign}
  \tilde{\Omega}^2_1 = {\cal M}(\epsilon_1)\left[\alpha-1- {\cal P}(\epsilon_1,\epsilon_2)\right].
  \label{eq:omega1bis}
\end{flalign}
  
The rotation rate of the host is obtained from (\ref{eq:bernoulli2}) at $E_2$ where the pressure is zero. On this surface, we have
\begin{flalign}
  Z^2=b_2^2\left(1-\frac{R^2}{a_2^2}\right),
\end{flalign}
which can be injected in (\ref{eq:bernoulli2}). At the lowest order, the $A_i''$'s do not depend on $\lambda$. The term in $R^2$ in this expression must therefore vanish, which yields the rotation rate, namely
\begin{flalign}
  \label{eq:omega2}
  &\tilde{\Omega}_2^2= {\cal M}(\epsilon_2)\left[1-(\alpha-1) \left. f_1 {\cal P}\left(\epsilon_2,\epsilon'_1\right)\right|_{{\rm B}_2} \right].
\end{flalign}
Let us now express the pressure in the host at the interface $E_1$. Still using (\ref{eq:bernoulli2}) but onto $E_1$, which is assumed to be a surface of constant pressure, we see that the term in $R^2$ has to be null, again. The rotation rate is then given by
\begin{flalign}
\label{eq:omega2order0constantpressure}
\tilde{\Omega^2_2} = \left. \left(A''_1 - \bareone^2 A''_3\right)\right|_{E_1},
\end{flalign}
which must be identical to (\ref{eq:omega2}). It is clear from (\ref{eq:aprim}) and (\ref{eq:adoubleprim}) that $A''_1=A'_i$ for $\lambda=0$. As a consequence, $\Omega^2_2=\Omega^2_1$. Rotation is therefore global. Since  all terms in $R^2$ are neutralized by the centrifugal potentials, and due to (\ref{eq:pbalance}), the pressure is a constant all along the interface $E_1$, which validates the initial assumption. The interface is therefore a surface of constant effective (gravitational and centrifugal) potential. This is called a ``type-C solution'' in the following. So, in the conditions of the approximation where $|c| \ll 1$, {\it in a heterogeneous systems made of two homogeneous components separated by spheroidal surfaces and rotating at the same rate, the pressure at the common interface is a constant.} We see by equating (\ref{eq:omega1bis}) to (\ref{eq:omega2}) that the necessary condition for global rotation is that the function 
\begin{flalign}
  \nonumber
 {\cal M}(\epsilon_2)&\left[1-(\alpha-1) \left. f_1 {\cal P}\left(\epsilon_2,\epsilon'_1\right)\right|_{{\rm B}_2} \right]\\& -{\cal M}(\epsilon_1)\left[\alpha-1- {\cal P}(\epsilon_1,\epsilon_2)\right] \equiv g(\epsilon_1,\epsilon_2,q,\alpha), 
  \label{eq:g}
\end{flalign}
admits at least one zero over the range of interest. Solving the equation $g=0$ requires in principle a full survey of the parameter space, which is $4$-dimensional, but it happens that $\alpha$ in (\ref{eq:g}) is easily accessible from the other parameters as it is observed in the confocal case \citep[e.g.][]{mmc83}, namely
\begin{flalign}
   \nonumber
 \alpha&= 1 + \frac{{\cal M}(\epsilon_2)+{\cal M}(\epsilon_1){\cal P}(\epsilon_1,\epsilon_2)}{{\cal M}(\epsilon_1)+{\cal M}(\epsilon_2) \left. f_1 {\cal P}\left(\epsilon_2,\epsilon'_1\right)\right|_{{\rm B}_2} },\\
 & \equiv \alphac.
  \label{eq:mdjump}
\end{flalign}
At order one in $c$, the rotation of the embedded spheroid is unchanged, as quoted, but $x$ varies slightly from point A$_1$ to point B$_1$ along the interface, which modifies $\Omega_2$. For the host, we use (\ref{eq:bernoulli2}) at these two end points, and we find
\begin{flalign}
- \frac{1}{2}\Omega_2^2 a_2^2 + \Psi_2|_{{\rm B}_2}- \Psi_2|_{{\rm A}_2}= 0,
  \label{eq:omega2-SCF}
\end{flalign}
 which can be put in a dimensionless form from (\ref{eq:psiint2}) and (\ref{eq:tildeomega}). In these conditions, it can then be shown that (\ref{eq:omega2}), (\ref{eq:g}) and (\ref{eq:mdjump}) keep the same form provided the quantity ${\cal M}(\epsilon_2)\left. f_1 {\cal P}\left(\epsilon_2,\epsilon'_1\right)\right|_{{\rm B}_2}$ is replaced by
\begin{flalign}
\label{eq:omega2correction}
{\cal M}(\epsilon_2) \left[\left. f_1 {\cal P}\left(\epsilon_2,\epsilon'_1\right)\right|_{{\rm B}_2} +  \underbrace{\left. f_1 {\cal C}(\epsilon_2,\epsilon'_1)\right|_{{\rm A}_2}^{{\rm B}_2}}_{\text{first-order correction}}\right],
\end{flalign}
where
\begin{flalign}
\label{eq:dq}
{\cal M}(\epsilon) f_1 {\cal C}(\epsilon,\epsilon')= f_1 \left[ A_0(\epsilon')x - (1- \epsilon^2) A_3(\epsilon') \right],
\end{flalign}
which is therefore to be evaluated at points A$_2$ and B$_2$. This correction represents the deviation to confocality and it vanishes when $c=0$.\\

\begin{figure}
\includegraphics[width=8.5cm,bb=90 60 514 427,clip==]{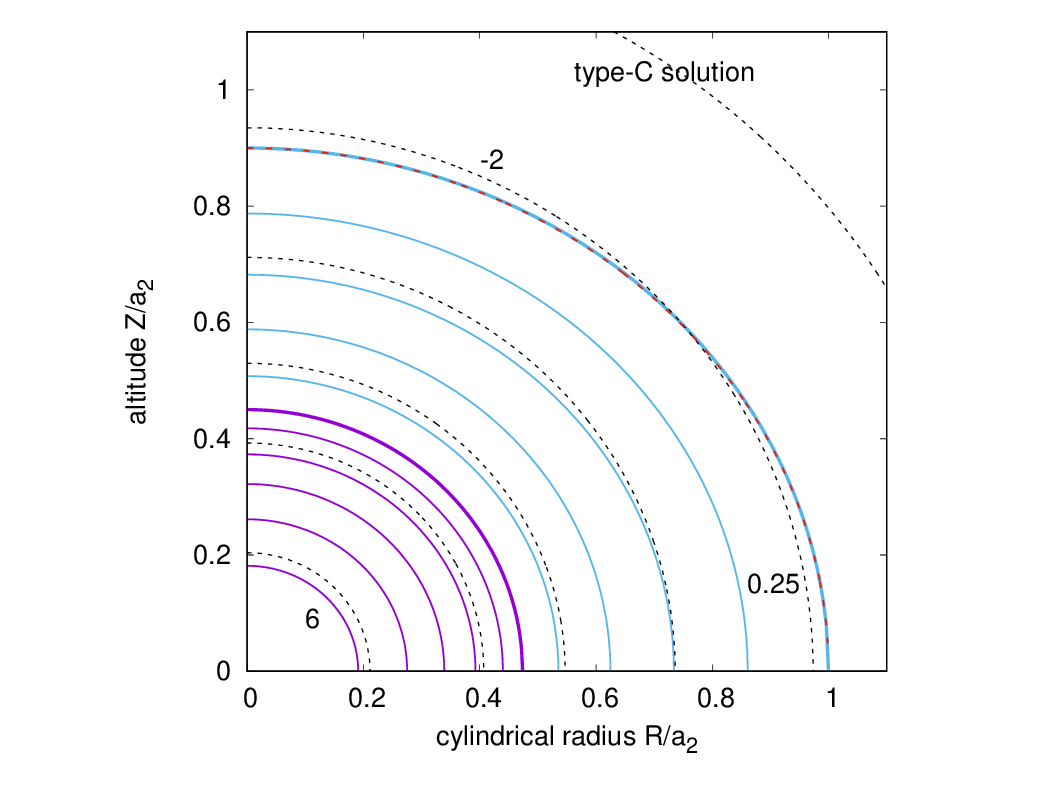}
\caption{The pressure in the embedded ellipsoid ({\it purple lines}) and in the host ({\it cyan lines}), the gravitational potential ({\it dotted lines}) and the spheroidal surfaces $E_1$ and $E_2$ ({\it bold lines}) for a type-C solution (configuration A); see Tab. \ref{tab:resultsc} (columns 4 and 5). Contour levels : step size $\delta p_2 =0.25$, $\delta p_1 =1$, $\delta \Psi=0.5$ and $p_2=0$ ({\it red dashed line}); see note \ref{note}.}
\label{fig:typec.ps}
\end{figure}

\begin{table}
  \centering
  \begin{tabular}{lrrrr}
\multicolumn{5}{c}{configuration A}  \\ \hline
    & this work    & {\tt DROP}$^\dagger$   & this work & {\tt DROP}$^\dagger$\\
    & order $0$       &                     & order $1$\\
    $\baretwo$      & $\leftarrow  0.90$ & $\leftarrow  0.90$ & $\leftarrow  0.90$&$\leftarrow  0.90$\\
    $\bareone$      & $\leftarrow 0.95$  &$0.94633$ & $\leftarrow 0.95$ & $0.95008$\\
    $q \bareone$    & $\leftarrow  0.45$  &$\leftarrow  0.45$& $\leftarrow  0.45$  &$\leftarrow  0.45$\\\\

    $q$             & $0.47368$           &$0.47637$ & $0.47368$  & $0.47364$\\
    $V/a_2^3$ & $3.76991$ & $3.76162$ & $3.76991$ & $3.76248$\\
    $c$             & $-0.16812$          &$-0.16556$ & $-0.16812$&$-0.16816$ \\     
    $x|_{{\rm A}_2}$ & $0.83187$ &&$0.83187$\\
    $\epsilon_1'|_{{\rm A}_2}$ & $0.16216$&&$0.16216$\\
    $f_1|_{{\rm A}_2}$ & $0.13486$&&$0.13486$\\
     $\epsilon_1'|_{{\rm B}_2}$ & $0.14790$&&$0.14790$\\
    $f_1|_{{\rm B}_2}$ & $0.10209$&&$0.10209$\\\\
   $\alpha \equiv \alpha_C$ & $5.70737$ & $\leftarrow \alpha_C$ & $6.35575$ & $\leftarrow \alpha_C$\\
    $\pc/\pi G\rho_2^2 a_2^2$    & $6.92857$ & $5.75910$&   $6.92857$& $6.93562$\\ 
$p^*|_{E_1}/\pi G\rho_2^2 a_2^2$   & $1.21192$ & $1.12135$ &  $1.21192$& $1.21078$\\  
$\tilde{\Omega}_1^2$ & $0.10918$ & $0.11777$ & $0.12627$& $0.12684$ \\   
$\tilde{\Omega}_2^2$ & $0.10918$ & $0.11777$ & $0.12627$& $0.12684$ \\
$M/\rhoenv a_2^3$ & $5.76085$ & $5.77289$ & $6.03507$ & $6.02797$ \\
    $\nu_1$ & $0.41901$ & $0.42222$ & $0.44541$ & $0.44617$ \\\hline
\multicolumn{5}{l}{$\leftarrow$ input data}\\
\multicolumn{5}{l}{$^*$value on the polar axis}\\
\multicolumn{5}{l}{$^\dagger$SCF-method \citep{bh21}}\\
  \end{tabular} 
   \caption{Data associated with Fig. \ref{fig:typec.ps} for order $0$ (column 2) and for order $1$ (column 4) in the $c$-parameter. The results obtained from {\tt DROP}-code (with a numerical resolution of $\frac{1}{127}$ corresponding to $7$ multigrid levels on a square grid) are also given (columns 3 and 5, respectively). Most numbers are truncated (five significant digits). See also note \ref{note}.}
  \label{tab:resultsc}
\end{table}

\begin{figure}
\includegraphics[width=8.7cm,bb=50 80 410 282,clip==]{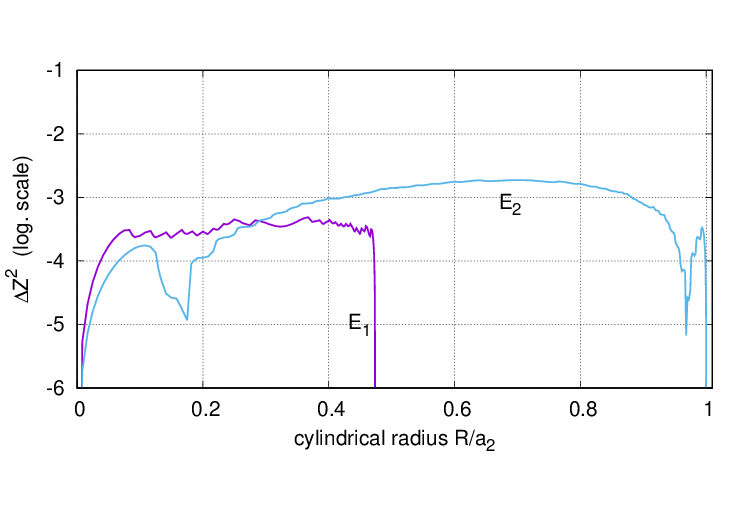}
\caption{Absolute deviation between the surfaces (meridional section) at equilibrium as computed from the SCF-method and the ellipses, for the embedded body ({\it purple}) and for the host ({\it cyan}), for configuration A; see also Fig. \ref{fig:typec.ps} and Tab. \ref{tab:resultsc}.}
\label{fig:devellipses.ps}
\end{figure}

Figure \ref{fig:typec.ps} displays an example of equilibrium in the form of contours levels for the pressure and for the gravitational potential\footnote{In the graphs and tables, the pressure is given in units of $\pi G\rho_2^2 a_2^2$ and the potential is in units of $\pi G \rho_2 a_2^2$\label{note}.} obtained for a canonical triplet $(q\bareone,\baretwo,q)$. The main input and output quantities are listed in Tab. \ref{tab:resultsc} for order $0$ (column 2) and for order $1$ (colmun 4). In this example, the confocal parameter is about $-0.17$, and the mass-density jumps are $\alphac \approx 5.7$ and $\approx 6.4$, respectively. By comparing the exact root of the second-order polynomial (\ref{eq:p2x}) with $x$ as given by (\ref{eq:xexpansionbis}), we can easily verify that the approximation is justified. This is corroborated by the results obtained with the {\tt DROP}-code (see Tab. \ref{tab:resultsc}, columns 3 and 5) that solves numerically the $2$-layer problem from the Self-Consistent-Field (SCF) method\footnote{This code includes an accurate determination of the bounding surfaces at each step of the SCF-cycle. In this process, the location of point B$_2$ is not known in advance (output).} \citep{bh21}. We show in Fig. \ref{fig:devellipses.ps} the absolute deviation between the ``true'' bounding surfaces and the ellipses $E_1$ and $E_2$, which is of the order of a few purcents. Equilibrium values (rotation rate, pressures, mass) are reproduced with a relative error of $5 \%$ typically at order 0, while this is of the order of $0.1 \%$ at order 1.\\

\begin{figure}
\includegraphics[width=8.7cm,bb=100 50 504 427,clip==]{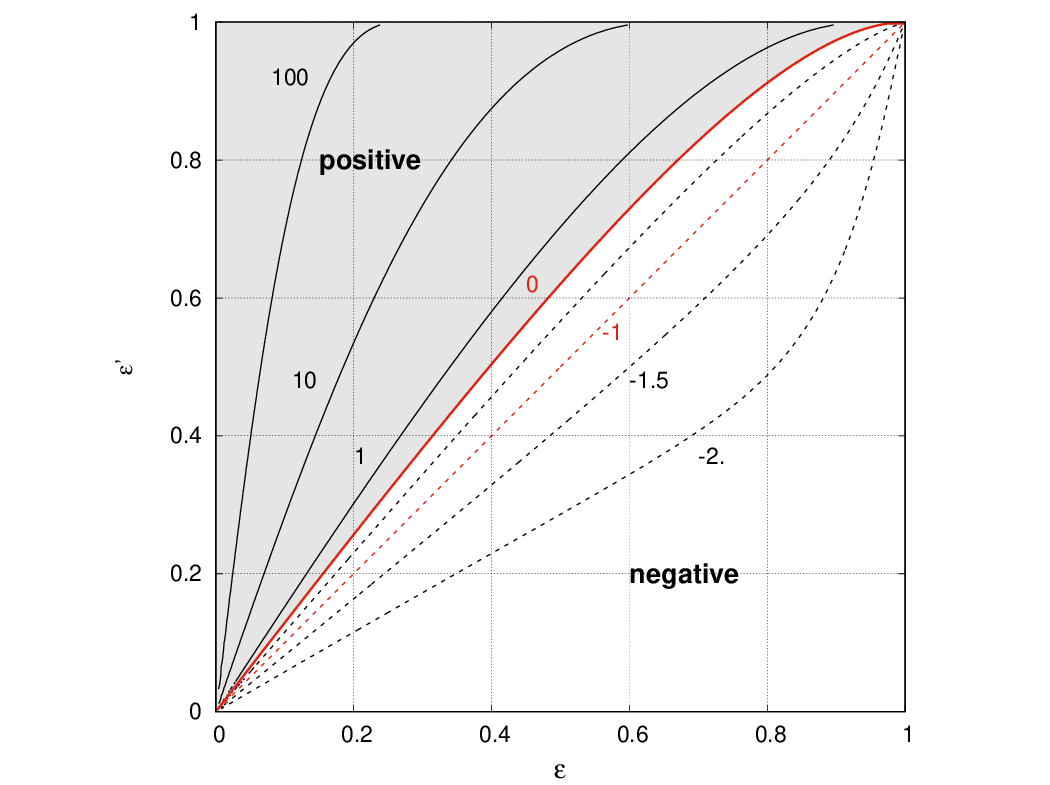}
\caption{Iso-contours of ${\cal P}(\epsilon,\epsilon')$. Positive values ({\it plain lines, shaded domain}; constant log. step $1$) and negative values ({\it dashed lines}; step size $0.5$) have been separated. Note that ${\cal P}(\epsilon,\epsilon)=-1$. See Tab. \ref{tab:solpiszero} for the solution of ${\cal P}(\epsilon,\epsilon')=0$.}
\label{fig:posxy.ps}
\end{figure}

\begin{figure}
  \includegraphics[width=8.7cm,bb=100 50 504 427,clip==]{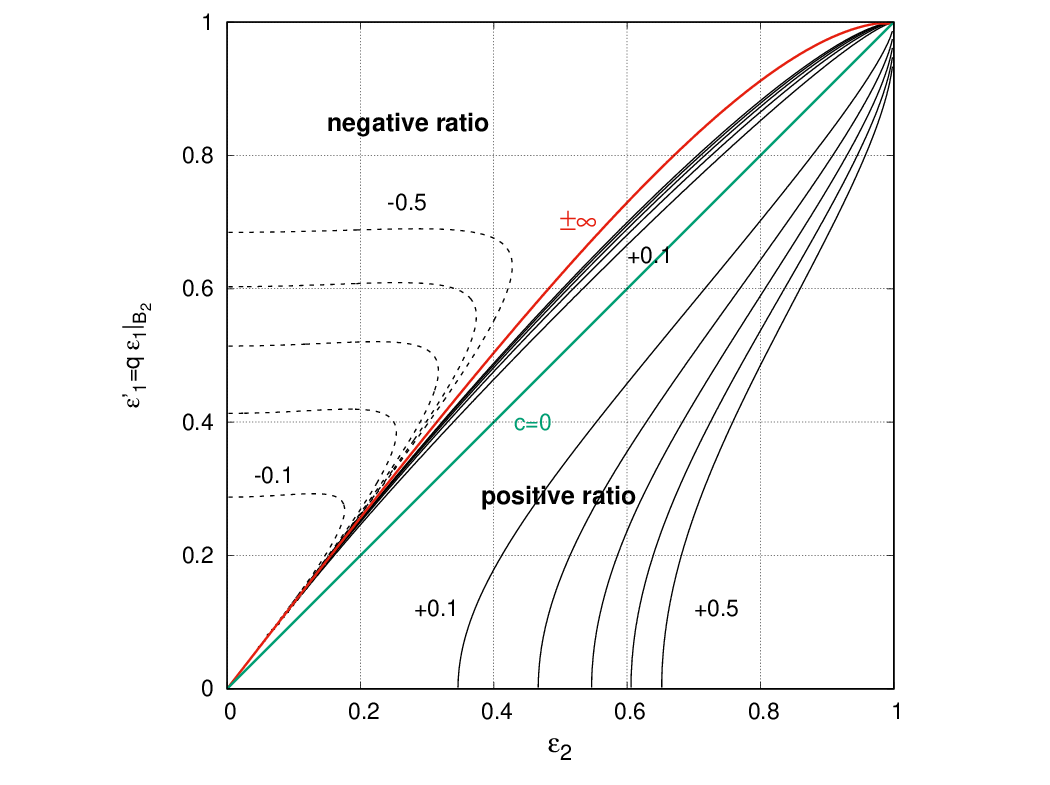}
\caption{Same as for Fig. \ref{fig:posxy.ps} but for the first-order correction relative divided by the leading term (constant step $0.1$), in the limit of $50\%$ in absolute. Also shown are the line where the first-order correction vanishes ({\it green line}) and the line where the leading term vanishes ({\it read line}); see also Fig. \ref{fig:posxy.ps} and  Tab. \ref{tab:solpiszero}.}
\label{fig:corxy.ps}
\end{figure}

\begin{table}
  \centering
  \begin{tabular}{llll}
   $\epsilon$  &$\epsilon'$ & $\epsilon$  &$\epsilon'$\\ \hline
    $0$            &$0$  & $0$            &$0$           \\   
    $0.050000$     &$0.064527$ &$0.038738$     &$0.050000$\\
    $0.100000$     &$0.128914$ &$0.077526$     &$0.100000$\\
    $0.150000$     &$0.193022$ &$0.116416$     &$0.150000$\\
    $0.200000$     &$0.256705$ &$0.155460$     &$0.200000$\\
    $0.250000$     &$0.319809$ &$0.194717$     &$0.250000$\\
    $0.300000$     &$0.382172$ &$0.234248$     &$0.300000$\\
    $0.350000$     &$0.443619$ &$0.274123$     &$0.350000$\\
    $0.400000$     &$0.503957$ &$0.314424$     &$0.400000$\\
    $0.450000$     &$0.562970$ &$0.355241$     &$0.450000$\\
    $0.500000$     &$0.620415$ &$0.396689$     &$0.500000$\\
    $0.550000$     &$0.676010$ &$0.438904$     &$0.550000$\\
    $0.600000$     &$0.729427$ &$0.482059$     &$0.600000$\\
    $0.650000$     &$0.780273$ &$0.526383$     &$0.650000$\\
    $0.700000$     &$0.828070$ &$0.572183$     &$0.700000$\\
    $0.750000$     &$0.872223$ &$0.619904$     &$0.750000$\\

    $0.800000$     &$0.911974$ &$0.670221$     &$0.800000$\\
    $0.850000$     &$0.946323$ &$0.724271$     &$0.850000$\\
    $0.900000$     &$0.973908$ &$0.784273$     &$0.900000$\\
    $0.950000$     &$0.992784$ &$0.855964$     &$0.950000$\\
    $1$            &$1$        &   $1$            &$1$       \\ \hline
\end{tabular}
 \caption{A six-digit solution of the equation ${\cal P}(\epsilon,\epsilon')=0$ for a regular sampling in $\epsilon$ (columns 1 and 2) and in $\epsilon'$ (columns 3 and 4); see also Fig. \ref{fig:posxy.ps}.}
  \label{tab:solpiszero}
\end{table}

\subsection{Singular cases. Condition of positivity}

There is a pending difficulty in the above relationship since $\alpha$ can diverge. This corresponds to a highly condensed embedded spheroid relative to the host, which situation can be associated with Roche systems \citep{jeans28}; see below. The singularity occurs for finite values of the parameters when the denominator in (\ref{eq:mdjump}) tends to $0$ (the sign of $\alphac$ changes), i.e. for
\begin{flalign}
{\cal M}(\epsilon_1) \rightarrow - {\cal M}(\epsilon_2) \left. f_1 {\cal P}\left(\epsilon_2,\epsilon'_1\right)\right|_{{\rm B}_2}.
\label{eq:singularalpha}
\end{flalign}
Figure \ref{fig:posxy.ps} displays ${\cal P}(\epsilon,\epsilon')$ in the form of contour levels. This quantity takes large positive values in the top-left part of the $(\epsilon,\epsilon')$-plane where $\epsilon' \gtrsim \epsilon$ roughly, and it takes small, negative values elsewhere. The ratio of the $1$st-order correction to the leading term is shown in Fig. \ref{fig:corxy.ps} in the $(\epsilon_2,q\epsilon_1  \equiv \epsilon'|_{{\rm B}_2})$-plane (this ratio depends on only these two parameters). It turns out that the correction is of small amplitude in relative in a wide part of the plane around the line $y=x$, except in three domains typically, namely: i) the vicinity of the line where ${\cal P}$ vanishes (see Tab. \ref{tab:solpiszero} for a sample of solutions of ${\cal P}(\epsilon,\epsilon')=0$), ii) when $\epsilon_2$ is close to unity (right part of the plot; the host resembles a flat disk), and iii) when $q\epsilon_1$ is close to unity (top part of the plot; the embedded ellipsoid is very flat, with a radius close to the radius of the host).

We see from (\ref{eq:mdjump}) that the magnitude and the sign of the mass-density jump remain hard to guess without considering numbers in the formula. The reason is that ${\cal P}(\epsilon_1,\epsilon_2)$ and ${\cal P}(\epsilon_2,q\epsilon_1)$ take opposite signs in the same domain of the $(\epsilon_2,\epsilon_1)$-plane. Besides, ${\cal M}$ is not a monotonic function of the ellipticity $\epsilon$, with a maximum value at $\epsilon \approx 0.930$. In fact, for  $\epsilon_1 < \epsilon_2$, we have ${\cal P}(\epsilon_1,\epsilon_2) \gtrsim 0$ and the numerator in (\ref{eq:mdjump}) is therefore poisitve and dominated by ${\cal M}(\epsilon_2)$. In the same time, ${\cal P}(\epsilon_2,q\epsilon_1) \lesssim 0$, making the denominateur of small amplitude, and eventually negative (unless $q$ is small). The mass-density jump $\alphac$ is large in absolute. In order to make the denominator positive (and subsequently to get a positive mass-density jump), $q$ must be small enough. This is the case of spheroids with a low oblateness and a massive host. The denominator in (\ref{eq:mdjump}) reaches zero by increasing $q$, and it becomes negative, which leads to $\alphac <0$. Note that configurations with $\epsilon_1 \gtrsim \epsilon_2 \sim 1$ corresponding to flat ellispoidal surfaces can be generated. This happens in the decreasing part of the function ${\cal M}(\epsilon)$ when $\epsilon \rightarrow 1$), but this involves $\alpha$-values close to unity.

Physically relevant solutions must be such that $\Omega_1^2>0$. Since the two rates are equal and ${\cal M}(\epsilon) >0$, this inequality writes, from (\ref{eq:omega1bis}) and (\ref{eq:omega2})
\begin{subnumcases}{}
  \alphac-1 \ge {\cal P}(\epsilon_1,\epsilon_2) \label{eq:thresholdc_omega1}\\
 1-(\alpha-1) \left. f_1 {\cal P}\left(\epsilon_2,\epsilon'_1\right)\right|_{{\rm B}_2} \ge 0 \label{eq:thresholdc_omega2}
\end{subnumcases}
It follows from the first condition that $\Omega_1^2$ is {\it inconditionally positive in the domain where ${\cal P}(\epsilon_1,\epsilon_2) < 0$}. From Fig. \ref{fig:posxy.ps}, we see that this domain corresponds to $\epsilon_1 \gtrsim \epsilon_2$. Such a condition, however, leads to small positive values, or even negative values, of $\alphac$. In contrast,  from (\ref{eq:thresholdc_omega2}), $\Omega_2^2$ is {\it inconditionally positive in the domain where ${\cal P}(\epsilon_2,\left. \epsilon'_1\right|_{{\rm B}_2}) < 0$, i.e. for $\epsilon_1 \lesssim \epsilon_2$ from  Fig. \ref{fig:posxy.ps}. This is precisely a situation that favours  large, positive values of the mass-density jump, which also makes (\ref{eq:thresholdc_omega1}) easily fulfilled}. We conclude that the most favorable conditions for the existence of nested figures of equilibrium in global rotation (type-C solutions) are met for $\epsilon_1^2 \lesssim \epsilon_2^2 \ll 1$ and small $q$-values.\\

\subsection{Special cases}

\noindent {\bf Confocality}. This geometry is met for $c=0$. In this case, the $1$st-order correction is zero, ${\cal P}(\epsilon_2,q\epsilon_1)=-1$ and (\ref{eq:mdjump}) reads
\begin{flalign}
 \alpha&= 1 + \frac{{\cal M}(q\epsilon_1)+{\cal M}(\epsilon_1){\cal P}(\epsilon_1,q\epsilon_1)}{{\cal M}(\epsilon_1)-\frac{{\cal M}(q\epsilon_1) q^3\bareone}{\sqrt{1-q^2\epsilon_1^2}}}.
  \label{eq:mdjumpconf}
\end{flalign}
A quick scan at the full domain $(\epsilon_1,q) \in [0,1]^2$ shows that $\alpha < 1$. Confocal states are exact solutions \citep{poincare88}, but the host must be more dense than the embedded spheroid, which is a highly unstable situation. This agrees with known results \citep{hamy90,mmc83}, and this is still true in the conditions of the actual approximation where $|c| \ll 1$: {\it a heterogeneous body made of two homogeneous components separated by confocal spheroids cannot be in global rotation (unless a density inversion)}.\\
 
\begin{figure}
\includegraphics[width=8.7cm,bb=100 50 514 427,clip==]{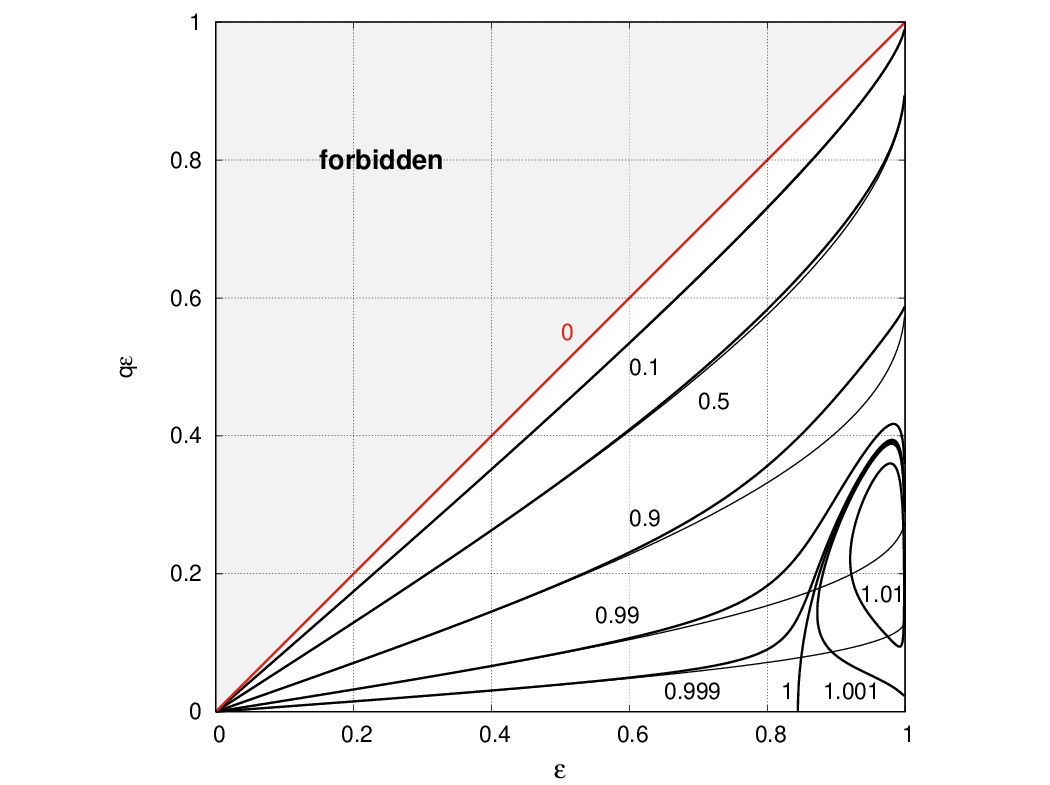}
\caption{Levels of contour for the function $h(\epsilon,q)$ in the $(\epsilon,q\epsilon)$-plane, at order $0$ ({\it thin lines}) and at order $1$ ({\it bold lines}).}
\label{fig:hxy.ps}
\end{figure}

\noindent {\bf Coellipticity}. For $\epsilon_1=\epsilon_2 \equiv \epsilon$, we have ${\cal P}(\epsilon,\epsilon)=-1$, and so, from (\ref{eq:g})
\begin{flalign}
  g(\epsilon,\epsilon,q,\alpha)=&-(\alpha-1) {\cal M}(\epsilon)h(\epsilon,q),
\end{flalign}
where
\begin{flalign}
  & h(\epsilon,q)=1+\left. f_1 {\cal P}\left(\epsilon,\epsilon'\right)\right|_{{\rm B}_2} +  \left. f_1 {\cal C}(\epsilon,\epsilon')\right|_{{\rm A}_2}^{{\rm B}_2}\\
  \nonumber
  &\qquad \qquad  =1+ q^3 \left\{ \frac{\bar{\epsilon} {\cal P}(\epsilon,q\epsilon)}{\sqrt{1-q^2 \epsilon^2}} + \frac{\bar{\epsilon} {\cal C}(\epsilon,q \epsilon)}{\sqrt{1-q^2 \epsilon^2}}   \right. \\
  \nonumber
  & \qquad \qquad\qquad  - \left. \frac{1}{[1 -\epsilon^2(1-q^2)]}{\cal C}\left(\epsilon,\frac{q \epsilon}{\sqrt{1 - \epsilon^2(1-q^2)}}\right)  \right\}.
\end{flalign}
If we exclude $\epsilon=0$ and $q=1$ as trivial solutions, $h$ keeps the same sign and does not vanish inside the relevant range $(\epsilon,q) \in [0,1]^2$, as Fig. \ref{fig:hxy.ps} proves. This is true at orders $0$ and $1$ in the $c$-parameter, and it does not depend on $\alpha$. Again, this agrees with \cite{hamy89}, and this is even true in the conditions of the approximation where $|c| \ll 1$: {\it a heterogeneous body made of two homogeneous components separated by similar spheroids cannot be in global rotation}.\\

\noindent {\bf The Maclaurin solution}. For $\alpha \rightarrow 1$, the density of the host and the density of the embedded spheroid become equal. We have respectively from ({\ref{eq:omega1bis}) and (\ref{eq:omega2})
\begin{subnumcases}{}
\tilde{\Omega}_1^2 \rightarrow - {\cal M}(\epsilon_1) {\cal P}(\epsilon_1,\epsilon_2),\\
\tilde{\Omega}_2^2 \rightarrow  {\cal M}(\epsilon_2).
\end{subnumcases}
As $\lim_{\epsilon_1 \rightarrow \epsilon_2} {\cal P}(\epsilon_1,\epsilon_2)=-1$, the two rotation rates merge only if $\epsilon_1 \rightarrow \epsilon_2$}. This result does not depend on $q$. The two components are indistinguishable and rotate in a synchroneous manner: this is basically a single object (Maclaurin) spheroid.\\

\begin{figure}
\includegraphics[width=8.7cm,bb=60 40 554 427,clip==]{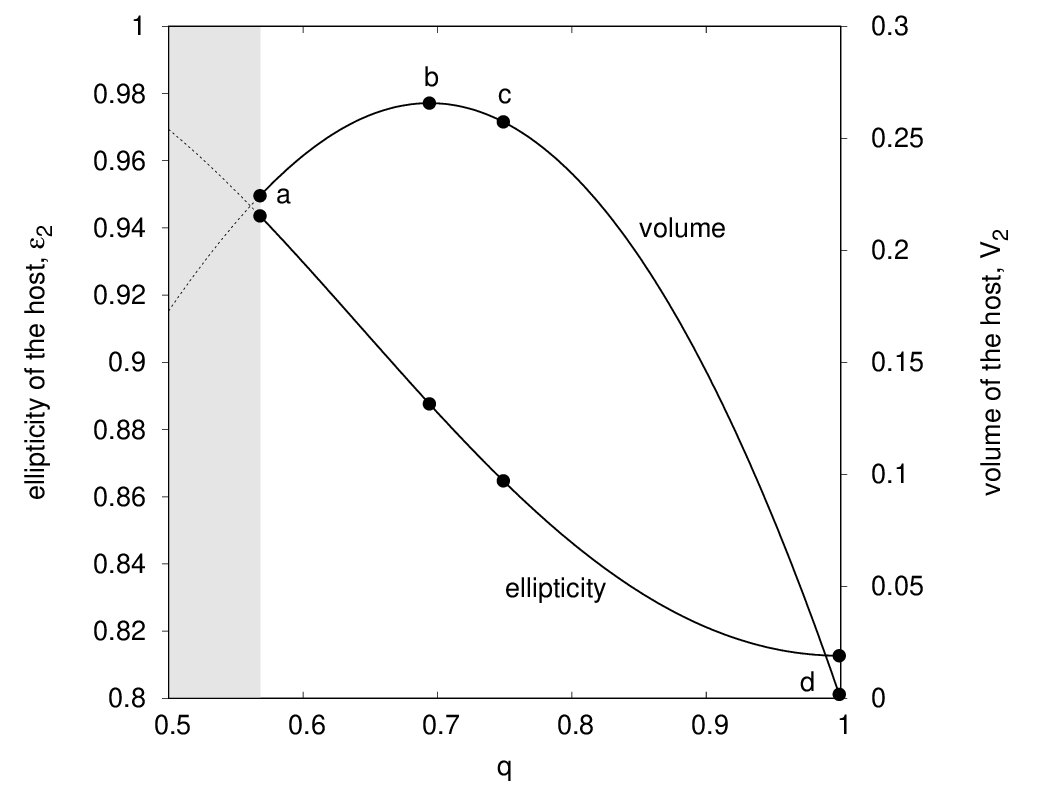}
\includegraphics[width=8.3cm,bb=0 0 1347 463,clip==]{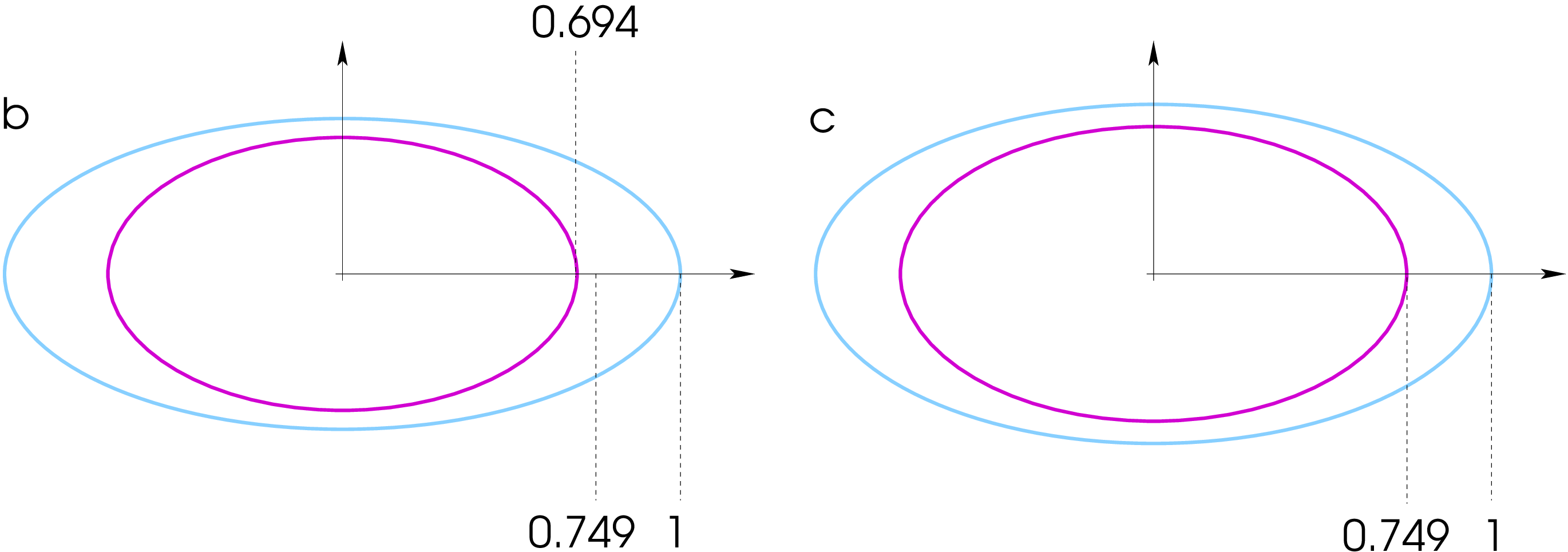}
\caption{Ellipticity of the host $\ellenv$ ({\it left axis}) and volume $V_2$ ({\it right axis}) versus the fractional radius $q$ for $\epsilon_1 \approx 0.812670$ ({\it top panel}), which corresponds to the transition towards the Jacobi ellispoidal sequence. For $q \lesssim 0.568$, the immersion condition is not fulfilled ({\it grey zone}). Four cases are highlighted ({\it black dots}): (a) the host has the largest equatorial extension and has the same polar radius as the embedded spheroid for $q \approx 0.568$ and $\bar{\epsilon_2} = \bar{\epsilon_1}  \approx 0.331$, (b) the volume of the host is maximum for $q \approx 0.694$ and $\epsilon_2 \approx 0.888$, (c) the solution is the one given by Jeans (1928) where $q\approx 0.749$ and $\epsilon_2 \approx 0.865$, (d) the host has null extension and coincides with the surface of the embedded spheroid for $q=1$ and $\epsilon_2=\epsilon_1$. The configurations associated with solutions (b) and (c) ({\it bottom panels}) compare successfully with Jeans' result (see Fig. 42 in his publication).}
\label{fig:jacobi_roche.ps}
\end{figure}

\noindent {\bf Generalized Roche systems}. Another interesting situation is met for $\alpha \rightarrow \infty$. In this case, the host is a rarefied medium compared to the embedded spheroid. From (\ref{eq:omega1bis}), we have $\tilde{\Omega}_1^2 \rightarrow \alpha {\cal M}(\epsilon_1)$, which is identical to (\ref{eq:omega2ref}). The embedded spheroid rotates by itself and carries away the host which has a negligible contribution to gravity. Actually, from (\ref{eq:omega2}), we get $\tilde{\Omega}_2^2 \rightarrow  \left. - \alpha {\cal M}(\ellenv) f_1 {\cal P}\left(\epsilon_2,\epsilon'_1\right)\right|_{{\rm B}_2}$, and by using (\ref{eq:singularalpha}), we recover $\tilde{\Omega}_2^2 \rightarrow {\cal M}(\ellcor)\alpha \equiv \tilde{\Omega}_1^2$. We can expand the expression for $\tilde{\Omega}_2^2$ in the limit $\epsilon_1 \rightarrow 0$. We have $A_0(0)=2$, $A_1(0)=A_3(0)=\frac{2}{3}$ and so we find
\begin{flalign}
\tilde{\Omega}_2^2 \approx -  \frac{4}{3} \alpha q^3 \left(1-\frac{1}{\baretwo}\right),
\label{eq:omega2roche}
\end{flalign}
where we have included the first-order correction. If we now express the rotation rate at the surface of the host (point B$_2$ of the equator; see Fig. \ref{fig:configsVCbis.eps}), as it is imposed by the embedded spheroid (now reduced to a point mass), we get $\tilde{\Omega}_1^2 = M/2 \pi \rho_2 a_2^3 \approx \frac{2}{3} \alpha q^3$, which is equal to (\ref{eq:omega2roche}) for $\baretwo=\frac{2}{3}$. This value is in agreement with Roche's model, although the true surface is not an ellipse \citep[e.g.][]{maeder2009}. This calculus can in principle be repeated for any value of $\epsilon_1$, in which case the equation $\tilde{\Omega}_2=\tilde{\Omega}_1$, if exists, yields a relationship between $q$ and $\epsilon_2$. As done in \cite{jeans28}, we have considered the ellipticity of the embedded (Maclaurin) spheroid at the bifurcation point towards the Jacobi sequence, namely $\epsilon_1 \approx 0.812670$ where $\tilde{\Omega}_1^2 \approx 0.18711$ \citep{chandra69}. The numerical solution $\epsilon_2(q)$ is plotted in Fig. \ref{fig:jacobi_roche.ps}. It fulfills (\ref{eq:immersion}) for $q \gtrsim 0.568$. For the lowest value, the host has zero thickness at the pole and the largest equatorial extension, while for $q=1$, it has zero thickness all along $E_1$ (which is therefore confunded with $E_2$). The volume of the host goes through a maximum at $q \approx 0.694$, which is close to the exact estimate by \cite{jeans28}, $q \approx 0.749$.

\section{Type-V solutions : the pressure varies along the interface}
\label{sec:typev}
        
We get the second family of solution in a very similar way, by considering the host first. Clearly, (\ref{eq:omega2}) is still valid, with or without the $1$st-order correction; see (\ref{eq:omega2correction}). The interface pressure, as imposed by the host, is therefore deduced from (\ref{eq:bernoulli2}), and it must be the same as the pressure delivered by the embedded spheroid. The main difference with above comes from (\ref{eq:omega2order0constantpressure}), which is no more imposed. This just means that the interface pressure participates in the mechanical support like the gravitatinal term in the Bernoulli equation, and it varies quadratically with the radius along $E_1$. This is called a ``type-V solution'' in the following. By combining (\ref{eq:bernoulli1}), (\ref{eq:bernoulli2}) and (\ref{eq:pbalance}), we find
\begin{flalign}
    \tilde{\Omega}_2^2 =  \tilde{\Omega}_1^2 + (\alpha-1)\left\{\tilde{\Omega}_1^2  - \left[ A'_1 -(1-\epsilon_1^2)A_3' \right]\right\},
  \label{eq:vtype}
\end{flalign}
which yields $\Omega_1^2$. We see that, if the quantity inside the curly brackets is zero, whatever the mass density jump, then $\Omega^2_2=\Omega^2_1$: this is precisely the type-C solution; see (\ref{eq:omega1}). Type-C solutions form a therefore a subset of type-V solutions. The other important point is that {\it the two components can be in relative rotation only if $\alpha \ne 1$}.

\begin{figure}
\includegraphics[width=8.5cm,bb=100 60 514 427,clip==]{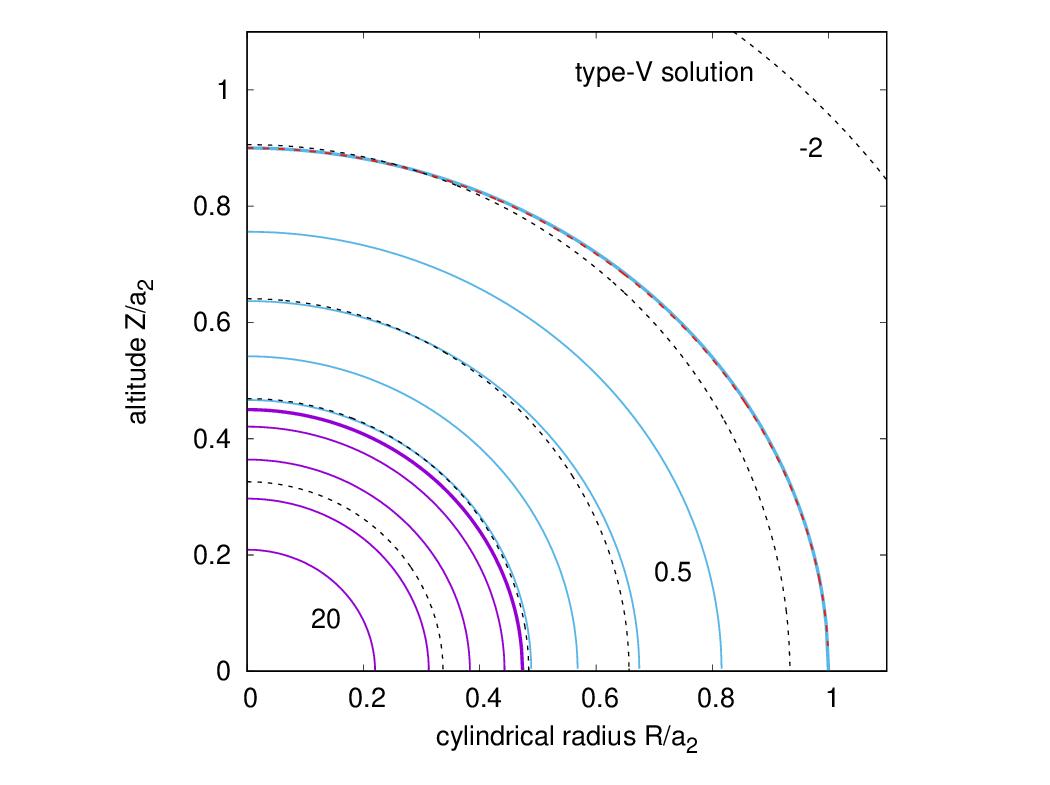}
\caption{Same legend as for Fig. \ref{fig:typec.ps}, but for a type-V solution with $\alpha=2 \alphac$; see Tab. \ref{tab:resultsv} (column 3; configuration B). Contour levels : step size $\delta p_2 =0.5$, $\delta p_1 =5$, $\delta \Psi=1$ and $p_2=0$ ({\it red, dashed lines}). See also note \ref{note}.}
\label{fig:typev.ps}
\end{figure}

\begin{figure}
\includegraphics[width=4.3cm,bb=210 60 424 427,clip==]{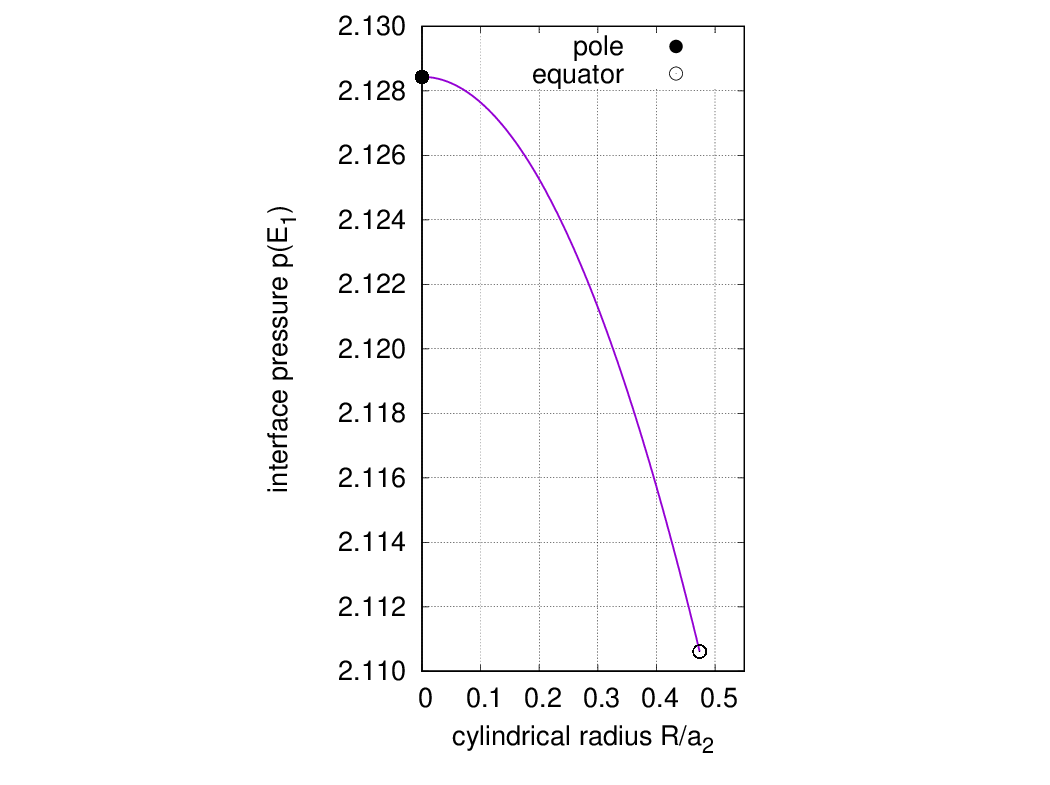}\includegraphics[width=4.3cm,bb=210 60 424 427,clip==]{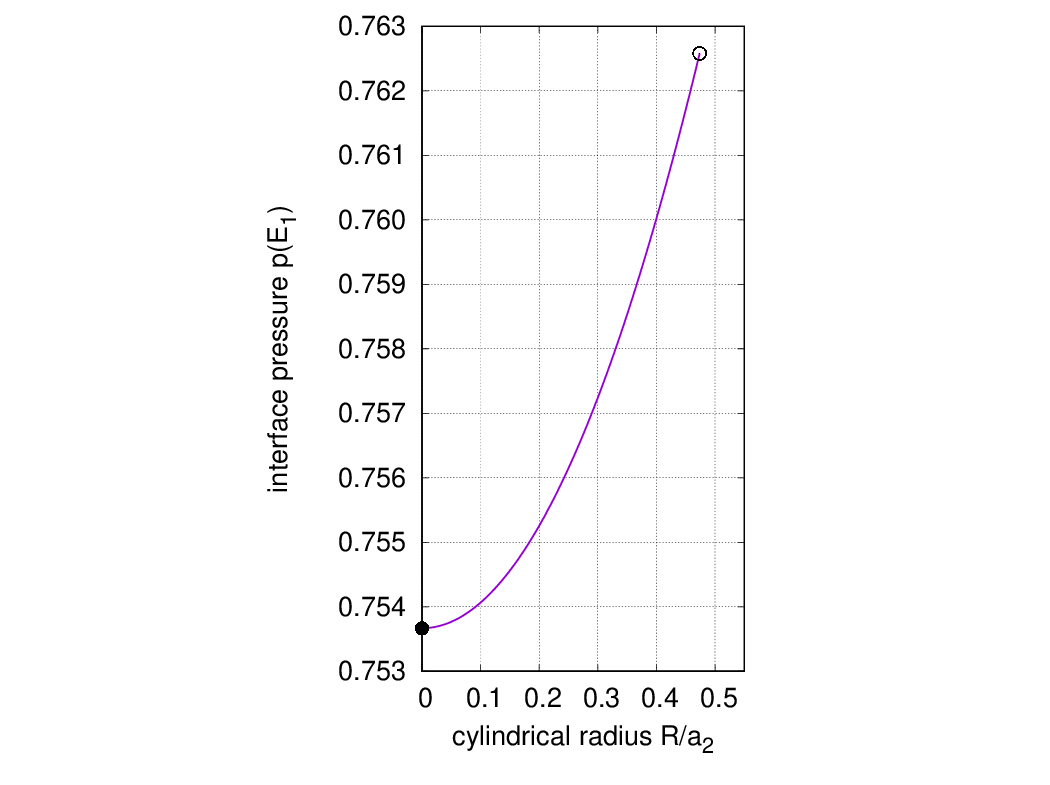}
\caption{Interface pressure (normalized, see note \ref{note}) versus the radius $R/a_2$ for configurations B ({\it left panel}) and C ({\it right panel}) reported in Tab. \ref{tab:resultsv}.}
\label{fig:pinterface.ps}
\end{figure}

Figure \ref{fig:typev.ps} displays an example of a type-V solution obtained for the same triplet as for Fig. \ref{fig:typec.ps}, but with $\alpha=2 \alphac$ (i.e. twice the value required by the type-C solution). The $1$st-order correction has been accounted for. The key quantities are given in Tab. \ref{tab:resultsv} (column 2). In this case, the rotation rate of the embedded spheroid is slightly larger than for the host. We can get a reverse situation if the mass-density jump is below the value corresponding to the type-C solution. We give in Tab. \ref{tab:resultsv} (column 4) the results obtained for $\alpha=\alphac/2$. Figure \ref{fig:pinterface.ps} shows the interface pressure as a function of the radius for these two examples. We notice that the variation from the pole to the equator is very weak, because the ellipticities of the two components are close to each other. Again, the comparison with the equilibrium states computed from the SCF-method  (see Tab. \ref{tab:resultsv}, columns 3 and 5) is very satisfactory as the relative deviations are of the order of $10^{-3}$.\\

\begin{table}
  \centering
  \begin{tabular}{lrrrr}
      & \multicolumn{2}{c}{configuration B} & \multicolumn{2}{c}{configuration C}\\\hline    
    & this work      & {\tt DROP}$^\dagger$ & this work & {\tt DROP}$^\dagger$\\
    $\baretwo$      & $\leftarrow  0.90$ & $\leftarrow 0.90$ & $\leftarrow  0.90$ &$\leftarrow  0.90$\\
    $\bareone$      & $\leftarrow 0.95$ & $0.95007$ & $\leftarrow 0.95$ & $0.94999$\\
    $q \bareone$    & $\leftarrow 0.45$ & $\leftarrow 0.45$  & $\leftarrow 0.45$ & $\leftarrow 0.45$\\\\
    $V/a_2^3$ & $3.76991$ & $3.75471$ & $3.76991$ & $3.76646$\\
    $q$             & $0.47368$         & $0.47364$ & $0.47368$ & $0.47368$\\
    $c$             & $-0.16812$        & $ -0.16815$ & $-0.16812$ & $-0.16811$\\\\     
   $\alpha$ & $\leftarrow  2\alpha_C$ & $\leftarrow  2\alpha_C$ & $\leftarrow  \frac{1}{2}\alpha_C$ & $\leftarrow  \frac{1}{2}\alpha_C$\\
    $\pc/\pi G\rho_2^2 a_2^2$                & $24.91929$ & $24.95266$ & $2.19229$ & $2.19346$\\ 
    $p^*|_{E_1}/\pi G\rho_2^2 a_2^2$           & $2.12843$  & $2.12529$ & $0.75366$ &  $0.75313$\\  
    $\tilde{\Omega}_1^2$ & $0.28756$  & $0.28847$ & $0.05499$ & $0.05534$\\   
    $\tilde{\Omega}_2^2$ & $0.21438$  & $0.21505$ & $0.08221$ & $0.08274$\\
    $M/\rhoenv a_2^3$    & $8.72318$  & $8.71163$ & $4.67840$ & $4.68772$\\
    $\nu_1$              & $0.61631$  & $0.61758$ & $0.28651$ & $0.28676$ \\\hline
  \multicolumn{4}{l}{$\leftarrow$ input data}\\
   \multicolumn{4}{l}{$^*$value on the polar axis}\\
 \multicolumn{4}{l}{$^\dagger$SCF-method \citep{bh21}}\\
   \end{tabular}
   \caption{Same legend as for Tab. \ref{tab:resultsc} but for two type-V solutions ($1$rst-order correction included); see also Figs. \ref{fig:typev.ps} and \ref{fig:pinterface.ps}a for configuration B. See also note \ref{note}.}
  \label{tab:resultsv}
\end{table}
 
\begin{figure}
\includegraphics[width=8.5cm,bb=100 60 514 427,clip==]{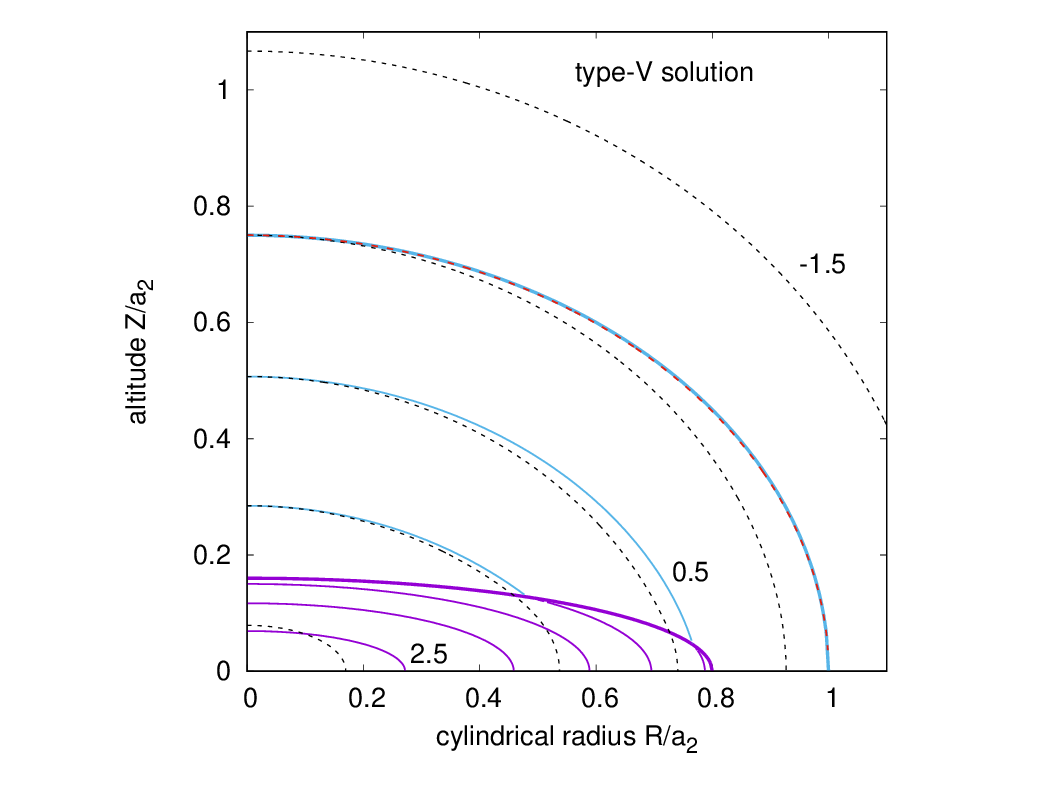}
\caption{A type-V solution for a flattened embedded spheroid inside a quasi-spherical host, and $\alpha=\alphac$; see Tab. \ref{tab:results2} (column 2; configuration D). Contour levels : step size $\delta p_2=\delta p_1 =0.5$, $\delta \Psi=0.5$, and $p_2=0$ ({\it red, dashed lines}). See also note \ref{note}.}
\label{fig:typev_diskinhalo.ps}
\end{figure}

\begin{figure}
\includegraphics[width=8.5cm,bb=100 60 514 427,clip==]{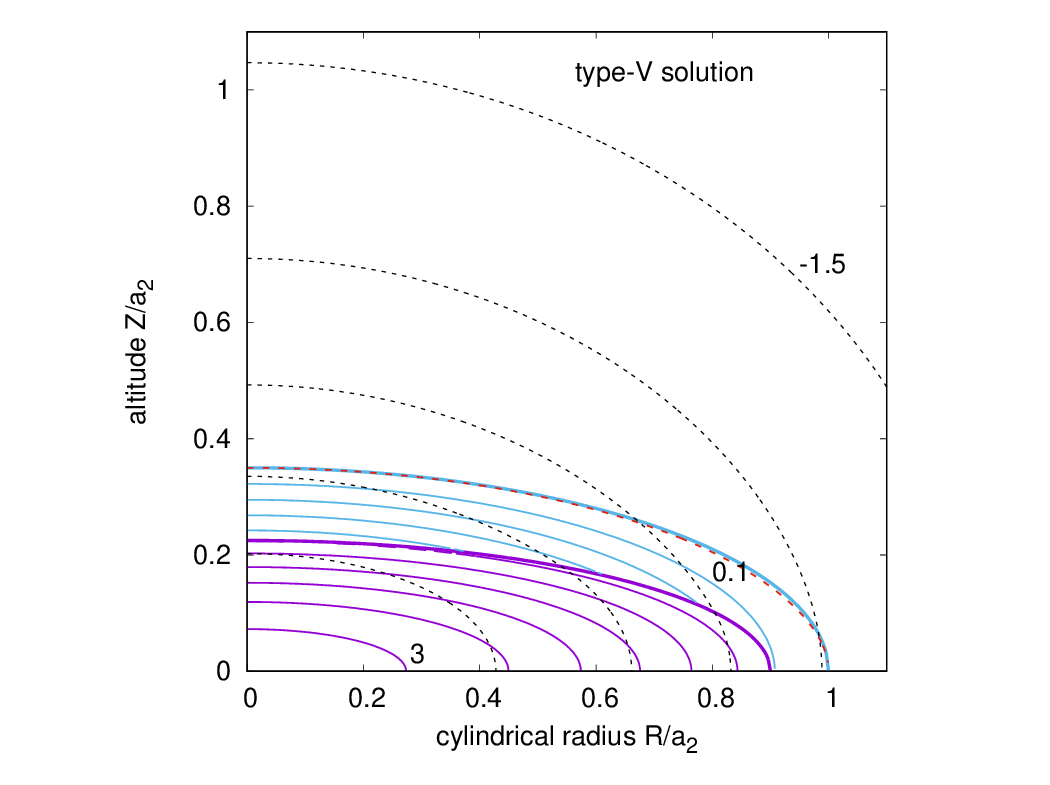}
\caption{A type-V solution for two flattened spheroids; see Tab. \ref{tab:results2} (column 3; configuration E) for the parameters. Contour levels : step size $\delta p_2 =0.1$, $\delta p_1 =0.5$, $\delta \Psi=1$, and $p_2=0$ ({\it red, dashed lines}). See also note \ref{note}.}
\label{fig:typev_diskindisk.ps}
\end{figure}

\begin{figure}
\includegraphics[width=8.5cm,bb=100 60 514 427,clip==]{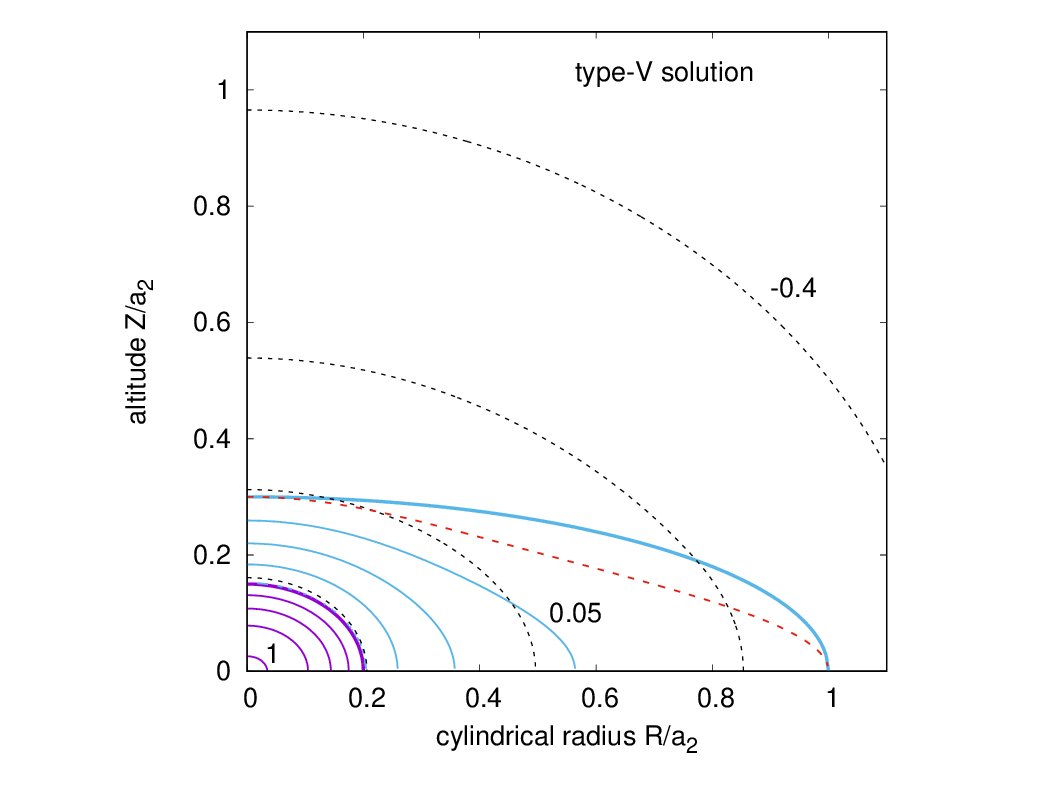}
\caption{A type-V solution corresponding to a weakly oblate spheroid inside a highly flattened host; see Tab. \ref{tab:results2} (column 4; configuration F) for the parameters. Contour levels : step size $\delta p_2 =0.05$, $\delta p_2 =0.2$, $\delta \Psi=0.2$ and $p_2=0$ ({\it red, dashed lines}). See also note \ref{note}.}
\label{fig:typev_starindisk.ps}
\end{figure}

It is interesting to see how the approximation behaves when the ellipticities are not ``small''. We show in Figs. \ref{fig:typev_diskinhalo.ps}, \ref{fig:typev_diskindisk.ps} and \ref{fig:typev_starindisk.ps} three typical type-V solutions obtained for the input parameters listed in Tab. \ref{tab:results2}. Again, the $1$st-order correction has been included in the calculations. We have conserved the same mass-density jump as above. The first case is a highly flattened embedded spheroid and a weakly oblate host. The $c$-parameter is still lower than unity, but now positive. It means that the host is less oblate than the confocal configuration would produce. The approximation is still very good (the line where the pressure naturally vanishes almost coincides with $E_2$). The rotation rate of the host is lower that for the embedded spheroid. The second case corresponds to two hilghly flattened bodies, again with a low $c$-parameter. The third example is a highly flattened host containing a moderately oblate, embedded body. The $c$-parameter is close to unity in absolute. The approximation of rigid rotation therefore fails. The variation of $\lambda$ is no more dominated by $\varpi^2$. This is visible in the figure since the line where $p_2=0$ and $E_2$ are no more confunded. Note that $\epsilon_1$ is beyond the threshold for dynamical stability (for a single body), and it would be interested to see the role of the host; see Sect. \ref{sec:conclusion}.\\

\begin{table}
  \centering
  \begin{tabular}{lrrr}
 & config. D & config. E & config. F \\\hline    
    $\bareone$      & $\leftarrow  0.2$   & $\leftarrow  0.25$   & $\leftarrow  0.75$ \\
    $\baretwo$      & $\leftarrow  0.75$  & $\leftarrow  0.35$    & $\leftarrow  0.3$ \\
    $q$             & $\leftarrow  0.8$   & $\leftarrow  0.9$   & $\leftarrow  0.2$ \\
    $\alpha$ & $\leftarrow  \alpha_C$ & $\leftarrow  \alpha_C$ & $\leftarrow  \alpha_C$ \\\\
    $q \bareone$    & $0.16000$ & $0.22500$  & $0.15000$\\
    $V/a_2^3$              &  $3.14159$ & $1.46607$ & $1.25663$ \\
    $c$                         & $0.17690$ & $-0.11812$ & $-0.89250$ \\\\
    $\pc/\pi G\rho_2^2 a_2^2$            & $2.77033$ & $3.29370$ & $1.02440$\\ 
    $p^*|_{E_1}/\pi G\rho_2^2 a_2^2$       & $1.32810$ & $0.46765$ & $0.20329$\\  
    $\tilde{\Omega}_1^2$     & $1.34745$ & $1.34636$ & $0.26435$ \\   
    $\tilde{\Omega}_2^2$     & $0.15973$ & $0.97739$ & $0.31436$ \\
    $M/\rho_2 a_2^3$              &  $5.43884$ & $5.55470$ & $1.39124$ \\
    $\nu_1$          &  $0.50124$ & $0.87350$ & $0.11481$ \\\hline
  \multicolumn{4}{l}{$\leftarrow$ input data}\\
   \multicolumn{4}{l}{$^*$value on the polar axis}\\
 \multicolumn{4}{l}{$^\dagger$SCF-method \citep{bh21}}\\
\end{tabular}
  \caption{Data for type-V solutions  ($1$rst-order correction included) corresponding to Figs.  \ref{fig:typev_diskinhalo.ps}, \ref{fig:typev_diskindisk.ps} and \ref{fig:typev_starindisk.ps}. See also note \ref{note}.}
  \label{tab:results2}
\end{table}

\subsection{Conditions of positivity}

Because $\alpha$ is free, type-V solutions are less contrained than type-C solutions. There are, however, still restrictions on the parameter sets leading to physically relevant solutions. The two rotation rates must be positive (double condition). Again, it is difficult to conclude since (\ref{eq:omega2}) and (\ref{eq:vtype}) are complicated funtions of $\epsilon_1$, $\epsilon_2$, $q$ and $\alpha$. If we omit the $1$st-order correction, the condition $\Omega_2^2  > 0$ writes
\begin{flalign}
  1-(\alpha-1) \frac{q^3\bareone}{\sqrt{1-q^2\epsilon_1^2}}{\cal P}\left(\epsilon_2,q\epsilon_1\right) \gtrsim 0,
  \label{eq:omega2cond}
\end{flalign}
and it is automatically verified as soon as ${\cal P}(\epsilon_2,q\epsilon_1)<0$ and (\ref{alphalargerthanone}) holds. We see from Fig. \ref{fig:posxy.ps} that this occurs for moderate/large values of $\epsilon_2$ and small/moderate values of $q\epsilon_1$. In the other part of the domain where ${\cal P}(\epsilon_2,q\epsilon_1)>0$, both $q$ and $\alpha$ play a critical role. The criterion can still be satisfied either with a value of $\alpha$ very close to unity or for a small $q$-parameter.

The second condition that must be examined simultaneously with (\ref{eq:omega2cond}) corresponds to $\Omega_1^2>0$. If we rewrite (\ref{eq:vtype}) as
\begin{flalign}
  \alpha \tilde{\Omega}_1^2 = \tilde{\Omega}_2^2 +(\alpha-1) {\cal M}(\epsilon_1) \left[\alpha-1- {\cal P}(\epsilon_1,\epsilon_2) \right],
  \label{eq:omega1v}
\end{flalign}
we see that the term inside the brackets can eventually be negative, but it must not exceed $\Omega_2^2$, in absolute. Again, for $\epsilon_1 < \epsilon_2$, we have ${\cal P}(\epsilon_1,\epsilon_2)<0$, which always ensures an equilibrium; see Fig. \ref{fig:posxy.ps}. For large, positive values of ${\cal P}(\epsilon_1,\epsilon_2)$, which occurs when the embedded boby is very flat with respect to the host, $\Omega_1^2$ can become negative. This situation can be ``neutralized'' in three ways: i) the mass-density jump is large enough in the sense $\alpha-1 > {\cal P}(\epsilon_1,\epsilon_2)$, ii) in constrast, $\alpha \rightarrow 1$ which decreases the term inside the brackets (we are close to global rotation in this case), and iii) ${\cal M}(\epsilon_1) \rightarrow 0$, which occurs for extreme values of $\epsilon_1$.\\

\subsection{Note about confocal and coelliptical configurations}

For $c=0$, (\ref{eq:omega2}) reads
\begin{flalign}
  \label{eq:omega2vtype}
  \tilde{\Omega}^2_2 = {\cal M}(q\epsilon_1)\left[1+(\alpha-1)\frac{q^3\bareone}{\sqrt{1-q^2 \epsilon_1^2}} \right],
\end{flalign}
which is clearly a positive quantity for $\alpha>1$, and $\Omega^2_1$ is easily deduced from (\ref{eq:omega1v}). Since $\epsilon_2 \le \epsilon_1$, ${\cal P}(\epsilon_1,\epsilon_2)$ is negative, meaning that $\Omega^2_1 >0$, but  $\Omega^2_1 \ne \Omega^2_2$. In agreement with \cite{mmc83}, and in the conditions of the approximation where $|c| \ll 1$, {\it the two homogeneous components of a heterogeneous body separated by confocal spheroids are necessarily in relative rotation}. Note that (\ref{eq:omega2vtype}) also writes
\begin{flalign}
  \Omega^2_2 = 2 \pi G \bar{\rho} {\cal M}(\epsilon_2),
\end{flalign}
where $\bar{\rho}$ is the mean density of the system.\\

By setting $\epsilon_1=\epsilon_2 \equiv \epsilon$ in (\ref{eq:vtype}), we find
\begin{flalign}
  \alpha \tilde{\Omega}_1^2 =  \tilde{\Omega}_2^2 +\alpha (\alpha-1) {\cal M}(\epsilon),
\end{flalign}
and (\ref{eq:omega2}) and (\ref{eq:omega2correction}) yield
\begin{flalign}
&\tilde{\Omega}^2_2 = {\cal M}(\epsilon)\\
\nonumber
&\quad\quad \times\left\{1-(\alpha-1)\left[\left. f_1 {\cal P}(\epsilon,q\epsilon)\right|_{{\rm B}_2} +  \left. f_1 {\cal C}(\epsilon,\epsilon')\right|_{{\rm A}_2}^{{\rm B}_2}\right]\right\}\\
\nonumber
&\quad= {\cal M}(\epsilon)\left\{1-(\alpha-1) \left[h(\epsilon,q)-1 \right] \right\}.
\end{flalign}
We see that the two rotation rates are identical only when $\alpha=1$ (single spheroid case), which is in agreement with the discussion in Sect. \ref{sec:typec}: {\it the two homogeneous components of a heterogeneous body separated by similar spheroids are necessarily in relative rotation}, in the conditions of the approximation where $|c| \ll 1$.

\section{Practical formula for the slow-rotation limit}
\label{sec:srl}

Cases with $\epsilon^2_2 \ll 1$ are of great interest for stars/planet interiors, as they correspond to the slow-rotation limit (the deformation with respect to sphericity is smaller than unity). It is generally admitted that interior layers are also characterized by small ellipticities, but there is no evidence that this occurs systematically, and this may depend on the process of formation, accretion and occasionally of differentiation of the entire body. If we set $\sqrt{k}=\epsilon_2/\epsilon_1$ (which is not necessarily small) and expand (\ref{eq:omega2}), (\ref{eq:dq}) and (\ref{eq:omega1v}) for small ellipticities, we find (see the Appendix \ref{sec:details} for more details)
\begin{flalign}
  \label{eq:slowrotationc2}
  &\tilde{\Omega}_2^2 \approx \frac{2}{15} \epsilon_1^2 \left\{ 2k + \frac{2}{7} k^2 \epsilon_1^2 \right.\\
  \nonumber
& \qquad   - (\alpha-1) q^3\left[3q^2-5k-\frac{3}{2}q^2\epsilon_1^2 \left(1+\frac{5}{14}q^2\right) \right.\\
  \nonumber
& \qquad \qquad\qquad \qquad \left. \left. + \frac{1}{15} k \epsilon_1^2(6q^2+5)-\frac{15}{4} k^2 \epsilon_1^2\right]\right\},
\end{flalign}  
for the host, and 
\begin{flalign}
  \label{eq:slowrotationc1}
  \alpha &\tilde{\Omega}_1^2 \approx \tilde{\Omega}_2^2 +\frac{2}{15}\epsilon_1^2 (\alpha-1)\\
  & \times \left[2(\alpha-1)\left(1+\frac{1}{7}\epsilon_1^2\right)+ 5-3k+2k\epsilon_1^2\left(1-\frac{6}{7}k\right)\right],
  \nonumber
\end{flalign}  
for the embedded spheroid. These formula include the $1$st-order correction. For type-C solutions, (\ref{eq:slowrotationc1}) directly yields
\begin{flalign}
  \label{eq:slowrotationglobalc1}
  \tilde{\Omega}_1^2 &\approx \frac{2}{15} \epsilon_1^2\\
  & \times \left[2(\alpha-1)\left(1+\frac{1}{7}\epsilon_1^2\right)+ 5-3k+2k\epsilon_1^2\left(1-\frac{6}{7}k\right)\right].
  \nonumber
\end{flalign}
By equating this expression to (\ref{eq:slowrotationc2}), we get the value of the mass-density jump; see (\ref{eq:slowrotc_alpha_1rstorder}) in the Appendix \ref{sec:details}. If $\alpha$ is significantly larger than unity or if a lower precision is sufficient, we can simplify more these formula and forget the $1$st-order correction. In these conditions, we get
\begin{subnumcases}{}
    \tilde{\Omega}_2^2 \approx \frac{2}{15}\epsilon_1^2 \left\{2k+ (\alpha-1) q^3(5k-3q^2)\right\},\qquad  \label{eq:slowrotationca}\\
    \alpha \tilde{\Omega}_1^2 \approx \tilde{\Omega}_2^2 +\frac{2}{15}\epsilon_1^2 (\alpha-1)\left[2\alpha-3(k-1)\right]. \label{eq:slowrotationcb}
    \end{subnumcases}

For type-C solutions, (\ref{eq:slowrotationcb}) yields
\begin{flalign}
  \tilde{\Omega}_1^2 \approx  \frac{2}{15} \epsilon_1^2\left[2\alpha-3(k-1)\right],
  \label{eq:slowrotc}
\end{flalign}
which is equal to (\ref{eq:slowrotationca}). Again, this gives the link between the mass-density jump, the size and the ellipticity of the embedded spheroid relative to the host, namely
\begin{flalign}
  \alpha \approx 1+ \frac{5(k-1)}{2+q^3(3q^2-5k)}.
  \label{eq:slowrotc_alpha}
\end{flalign}

\begin{figure}
\includegraphics[width=8.7cm,bb=100 50 514 427,clip==]{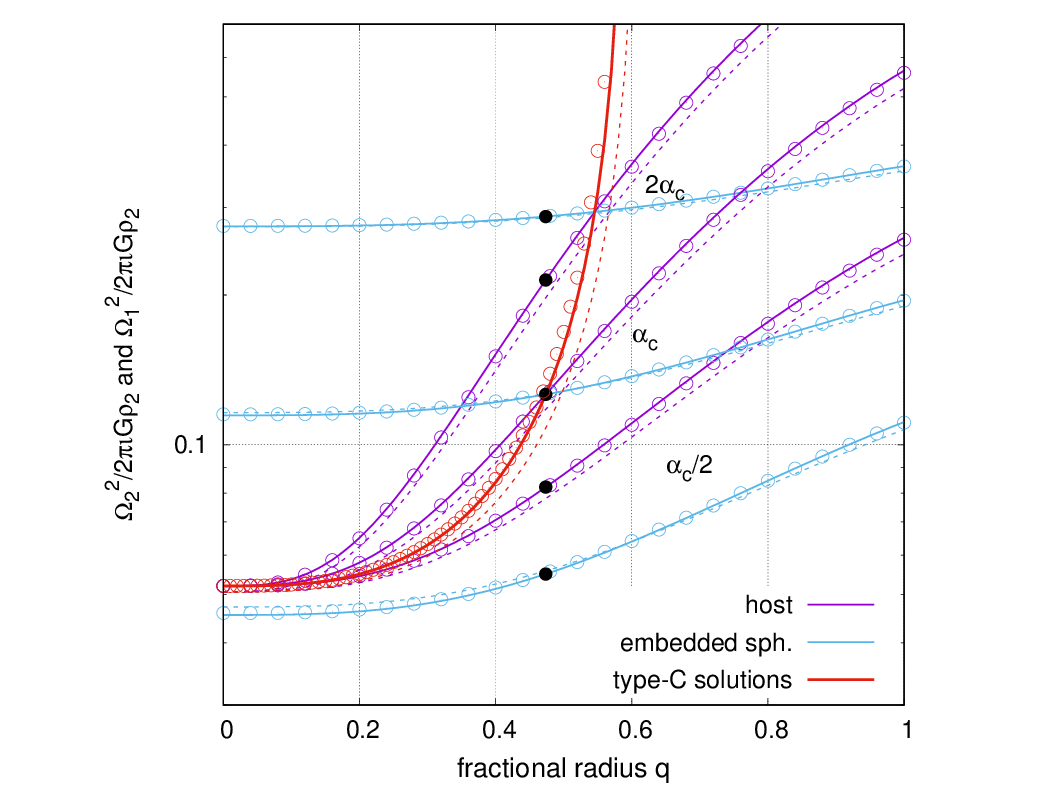}
\caption{Square of the rotation rate (normalized) for the embedded spheroid ({\it plain, cyan lines}) from (\ref{eq:omega1v}) and for the host ({\it plain, purple lines}) from (\ref{eq:omega2}) versus $q$ for $(\bareone,\baretwo)=(0.95,0.9)$, and for $3$ values of the mass-density jump $\alpha$. The case of global rotation (type-C solution) from (\ref{eq:slowrotationglobalc1}) is also shown ({\it red}). Also plotted is the leading term ({\it dashed lines}) of the expansion as given by (\ref{eq:slowrotationca}), (\ref{eq:slowrotationcb}) and (\ref{eq:slowrotc}), and the $1$st-order approximation ({\it open circles}) computed from (\ref{eq:slowrotationc2}), (\ref{eq:slowrotationc1}) and (\ref{eq:slowrotationglobalc1}); see Tabs. \ref{tab:resultsc} and \ref{tab:resultsv} for $q \approx 0.473$ ({\it black circles}).}
\label{fig:omega.ps}
\end{figure}

Figure \ref{fig:omega.ps} compares these approximations with the references (\ref{eq:omega2}) and (\ref{eq:omega1v}) for the pair $(\bareone,\baretwo)=(0.95,0.9)$ already considered in the preceeding sections, and three values of the $\alpha$-parameter. We see the excellent agreement between various formula. While the $1$st-order correction is, as expected, very precise, the leading term, alone, is already remarkably close to the reference, in particular for the embedded spheroid.

It is interesting to notice that (\ref{eq:slowrotc_alpha}) is compatible with the conclusions drawn in Sect. \ref{sec:typec}: global rotation is not possible for coelliptical configurations ($\alpha=1$ for $k=1$) and for confocal states ($\alpha < 1$ for $k=q^2$) as well.
 Another important point concerns the magnitude of $k$. By reversing (\ref{eq:slowrotc_alpha}), we find
\begin{flalign}
k \approx 1 + \frac{(\alpha-1)\left[2+q^3(3q^2-5)\right]}{5\left[1+ (\alpha-1)q^3\right]},
  \label{eq:slowrotc_k}
\end{flalign}
and we see that $k$ is larger than unity in the whole domain of interest $q \in [0,1]$ provided $\alpha >1$. This is in agreement with \cite{hamy89}, and in coherence with what is experimentally observed from the SCF-method \citep{bh21}. It follows that, in the conditions of the actual approximation where $|c| \ll 1$, {\it in a heterogeneous body made of two homogeneous, synchroneously rotating components separated by spheroidal surfaces, the embedded spheroid is necessarily more spherical (less oblate) than the host}. Note that $k \approx 1+\frac{2}{5}(\alpha-1)$ when $q^3 \rightarrow 0$ (the embedded spheroid has small size), while $k \approx 1$ for $q \rightarrow 1$ (the host has small size).

For type-V solutions, the ratio of the rotation rates is directly found from (\ref{eq:slowrotationca}) and (\ref{eq:slowrotationcb}), namely
\begin{flalign}
\frac{\Omega^2_1}{\Omega_2^2} \approx \frac{1}{\alpha}\left[ 1+ \frac{(\alpha-1)[2 \alpha-3(k-1)]}{2k+(\alpha-1)q^3(5k-3q^2)}\right].
  \label{eq:slowrotc_k}
\end{flalign}
Typically, this ratio is smaller than unity for large values of the $k$-parameter (the embedded spheroid is very close to spherical) and $q$ close to unity (the relative size/volume of the host is small).

\begin{table*}
  \begin{tabular}{lccl}
  \multicolumn{2}{c}{\bf input parameters} &  equation & comment \\ \hline
  ellipticity of $E_2$                 & $\epsilon_1 \in [0,1]$ \\
  ellipticity of $E_1$                 & $\epsilon_2 \in [0,1]$ \\
  fractional radius of the embedded spheroid & $q=a_1/a_2 \in [0,1]$\\
  mass-density jump                    & $\alpha \ge 1$ & & useless for type-C solutions\\
  \\
  \multicolumn{2}{c}{\bf intermediate data}\\\hline
  confocal parameter & $c$ & (\ref{eq:confocalc}) & $|c| \ll 1$ required \\
  various coefficients         & $A_0$, $A_1$ and $A_3$ & (\ref{eq:IA1A3}) \\
                               & $x$                   & (\ref{eq:xab}) & exact values at points A$_2$ and B$_2$\\
  & $f_1$ and $\epsilon'$  & (\ref{eq:feprimab})\\
  leading term & $\left. f_1 {\cal P}\left(\epsilon_2,\epsilon'_1\right)\right|_{{\rm B}_2}$ & (\ref{eq:pfunction}) \\
  $1$rst-order correction & $\left. f_1 {\cal C}(\epsilon_2,\epsilon'_1)\right|_{{\rm A}_2}^{{\rm B}_2}$ & (\ref{eq:omega2correction})\\
  \\
  \multicolumn{2}{c}{\bf pressure}     \\ \hline
  surface & $p|_{E_2}=0$ \\
  interface (pole value) & $p^*|_{E_1}$ & (\ref{eq:pinterface}) & see (\ref{eq:pbalance})\\
  central value & $\pc$ & (\ref{eq:pc})\\
  \\
  \multicolumn{2}{c}{\bf rotation rate (type-V solution)}     \\ \hline
  host              & $\Omega_2$ & (\ref{eq:omega2}) \\
  \qquad small ellipticities (order 0)& & (\ref{eq:slowrotationca}) & $\epsilon_1^2 \ll 1$ and $\epsilon_2^2 \ll 1$ \\
  \qquad small ellipticities (order 1)& & (\ref{eq:slowrotationc2}) & \\
  embedded spheroid & $\Omega_1$ & (\ref{eq:omega1v})\\
  \qquad small ellipticities (order 0) & & (\ref{eq:slowrotationcb}) & $\epsilon_1^2 \ll 1$ and $\epsilon_2^2 \ll 1$ \\
  \qquad small ellipticities (order 1) & & (\ref{eq:slowrotationc1}) & \\
  \\
  \multicolumn{2}{c}{\bf rotation rate  (type-C solution)}\\\hline
  both components & $\Omega_1 = \Omega_2$ & (\ref{eq:omega1bis}) or (\ref{eq:omega2})\\
  mass-density jump & $\alphac$ & (\ref{eq:mdjump})\\
  \\
  \qquad small ellipticities (order 0)   & & (\ref{eq:slowrotc}) and (\ref{eq:slowrotc_alpha}) for $\alphac$\\
  \qquad small ellipticities (order 1)   & & (\ref{eq:slowrotationc2}) or (\ref{eq:slowrotationc1}) and (\ref{eq:slowrotc_alpha_1rstorder}) for $\alphac$\\ \hline
  \end{tabular}
  \caption{Summary of useful formula. See the Appendix \ref{sec:f90} for a simple F90 program.}
  \label{tab:usefulprmula}
\end{table*}
    
\section{Conclusion and perspectives}
\label{sec:conclusion}

This article is a novel contribution to the theory of figures \citep{chandra69}. We have established the equilibrium conditions for a heterogeneous body made of two homogeneous components bounded by concentric and coaxial, spheroidal surfaces and in relative rotation. This special geometry offers a great mathematical simplification since the gravitational potential of spheroids is known in closed form. Regardless of the rotation laws, we can consider a wide range of flattenings that is difficult to reach through perturbative methods \citep{ch33,ca16}. Various collisional systems are concerned, like stars and planets and gaseous envelopes hosting protostars. Due to the hypothesis of incompressibility, however, the best targets for this study are rocky/icy planets surrounded by a solid/liquid envelope. A generalization of the present approach to the multi-layer case is proposed in \cite{h21b}.

The two-component problem depends on four parameters, three geometrical parameters (the ellipticities and the fractional size of the immersed body) and one thermodynamical parameter (the mass-density jump). This already renders the analytical treatement complicated. Except for a specific ambient pressure and for confocal configurations (Poincar\'e's theorem), there is no exact solution to the problem of nested spheroids compatible with rigid rotation. In the latter case, however, a mass-density inversion is necessary, which is highly improbable for stability reasons \citep{hamy90,moulton16,mmc83}.

As argued in \cite{hamy89}, states of rigid rotations are valuable in a first approximation only for small ellipticies. As shown here, the confocal parameter $c$ defined by (\ref{eq:confocalc}) enables to consider much more configurations than those accessible by assuming small ellipticities. This work can therefore be regarded as a prolongation of Hamy's approach. When $|c| \ll 1$, the problem admits typically two families of solutions, depending on the interface pressure. For type-C solutions, both components are in synchronous rotation (rotation is global), and the pressure is constant all along the common interface. In agreement with previous works, neither confocal configurations nor coelliptical configurations are permitted, and the ellipticity of the host must be larger than that of the embedded spheroid. For type-V solutions, the interface pressure varies quadratically with the cylindrical radius. The embedded spheroid and the host are necessarily in relative rotation, and this requires a mass-density jump. More configurations are possible with respect to type-C solutions. Confocal and coelliptical states are permitted. Depending on the fractional radius and on the mass-density jump, the host can rotate faster or slower than the embedded body.

As discussed, the conditions for the existence of nested spheroidal figures are preferentially fulfilled when the embedded spheroid is more spherical than the host, and for a small fractional radius, but it is clear that the criteria, namely (\ref{eq:immersion}), (\ref{alphalargerthanone}), (\ref{eq:thresholdc_omega1}), (\ref{eq:thresholdc_omega2}) and (\ref{eq:omega2cond}), must be carefully tested for each configuration by considering numbers in the formula. Both type-C and typ-V solutions have been validated through several examples. In particular, these compare sucessfully with the numerical approach based on the SCF-method  \citep{bh21} as long as the condition $|c| \ll 1$ is satisfied. In practice, the bounding surfaces deviate only slightly from pure ellipses, and the ``true'' gravitational potential differs from the expressions for the spheroids given in Sec. \ref{sec:hyp} only by a very small amount. In the case where the two spheroidal surfaces are close to spherical, which is appropriate for slowly-rotating stars and planets, we have derived a simple relationship for the rotation rate of each component, as  function of the main input parameters. In the case of global rotation, this yields a simple relationship between the mass-density jump, the fractional radius and the ellipticity ratio; see Sect. \ref{sec:srl}. We give in Tab. \ref{tab:usefulprmula} a summary of the most useful formula. A basic (non-optimized) program written in Fortran 90 is given in the Appendix \ref{sec:f90}.
    
\begin{figure}
\includegraphics[width=8.7cm,bb=90 50 524 427,clip==]{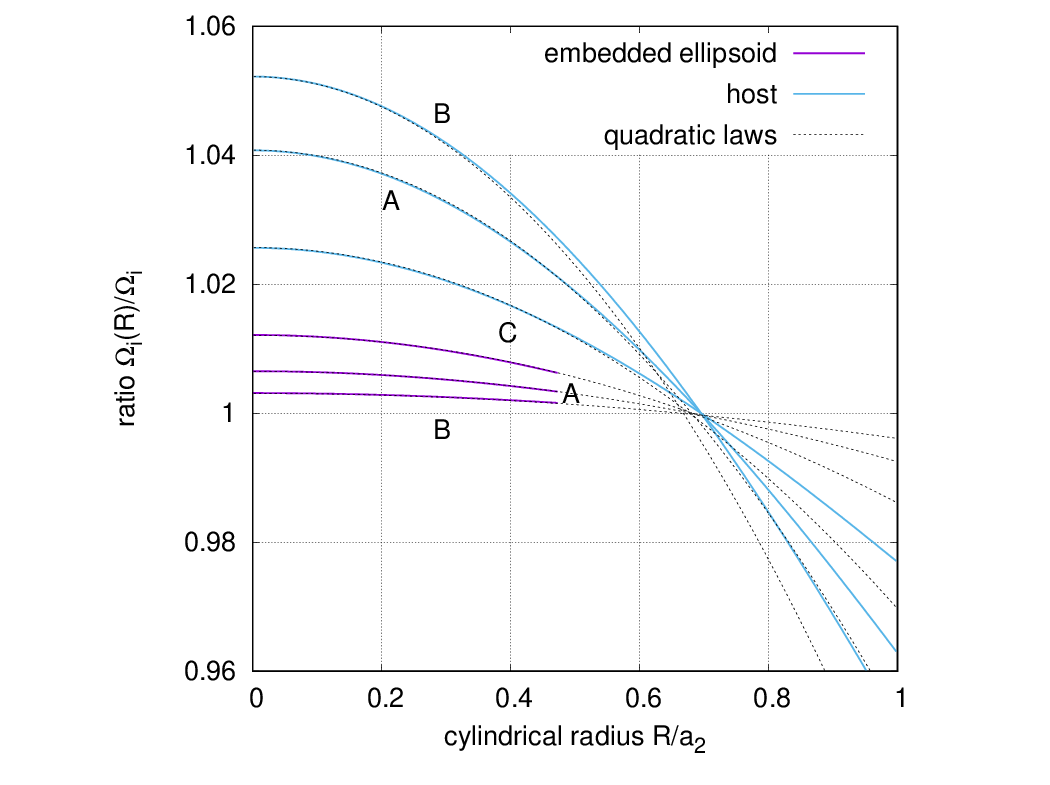}
\caption{Rotation profiles (normalized to the values for rigid rotation) for the exact solutions corresponding to the configurations A, B and C; see Tabs. \ref{tab:resultsc} and \ref{tab:resultsv}.}
\label{fig:phiABC.ps}
\end{figure}

A natural extension of this paper concerns the impact of the next term in the expansion of the gravitational potential. As shown, the ``coefficients'' $A''_i$''s in (\ref{eq:adoubleprim}) depend on $\lambda$, and subsequently on $R^2$ through (\ref{eq:p2x}). While $R^2$ appears as the leading term for small confocal parameters, higher powers are present and can be accounted for. This means to go beyond the assumption of rigid rotation, at least for the host. If
\begin{flalign}
\Phi_i(R)=-\int{\Omega_i^2(R)RdR},
  \label{eq:centrifugal}
\end{flalign}
denotes the centrifugal potential for component $i=\{1,2\}$, we have, as a generalization of (\ref{eq:bernoulli1}) and (\ref{eq:bernoulli2})
\begin{subnumcases}{}
    \frac{\pint}{\rho_1} + \Phi_1 - \pi G \rho_2 \left(A'_0 - A'_1 R^2 -A_3'Z^2 \right) =  \const, \qquad \label{eq:generala}\\
    \frac{\pext}{\rho_2} +\Phi_2 -\pi G \rho_2\left( A''_0 - A''_1 R^2 -A_3''Z^2 \right) =  \const'.\qquad \label{eq:generalb}
\end{subnumcases}{}
Clearly, $\Phi_2(R)$ is determined by evaluating (\ref{eq:generalb}) along $E_2$ where $p_2=0$, namely
\begin{flalign}
  \nonumber
  \frac{\Phi_2(R)}{\pi G \rho_2 a_2^2} =&   \frac{1}{a_2^2}\left[A''_0-A''_0|_{{\rm A}_2}\right]- (1-\epsilon_2^2) \left[A''_3-A''_3|_{{\rm A}_2}\right]\\
  &\quad - \left[A''_1-(1-\epsilon_2^2) A''_3\right]\varpi^2,
\end{flalign}
and the requirement of pressure balance onto $E_1$ enables to link $\Phi_1$ and $\Phi_2$ together. From (\ref{eq:pbalance}), (\ref{eq:generala}) and (\ref{eq:generalb}), we have 
\begin{flalign}
 \Phi_2 - \alpha \Phi_1 & - (\alpha-1) \Psi(E_1)-\const'+\alpha \, \const=0,
\end{flalign}
hence $\Phi_1(R)$. Then, the rotation profiles $\Omega_1(R)$ and $\Omega_2(R)$ are easily deduced from (\ref{eq:centrifugal}) by derivation. We show in Fig. \ref{fig:phiABC.ps} the rotation profile $\Omega_i(R)$ for the embedded spheroid and the host deduced from (\ref{eq:centrifugal}), (\ref{eq:generalb}) and (\ref{eq:generala}) for configurations A, B and C already considered (see Tabs. \ref{tab:resultsc} and \ref{tab:resultsv}). While the range of variation of $\Omega_i(R)$ is of the order of a few purcents (which validates the approximation), we clearly see an underlying quadratic law for $\Omega_i^2$, and even a quartic contribution for the host.

Another point that would merit some investigation concerns the dynamical stability of the system. A possible option is to reproduce the present analysis in the case of a two-components, triaxial body \citep[e.g.][]{mcm90}, and to compare the energies between the spheroidal and the ellipsoidal configurations. In this purpose, a dedicated code capable of solving the three dimensional problem numerically, for instance via the SCF-method, seems vital.

\section*{Data availability} All data are incorporated into the article.

\section*{Acknowledgements}

I am grateful to Pr. J. Tohline, Dr. B. Basillais and A. Meunier for advices and suggestions on the article. We thank the anonymous referee for a very detailed examination of the paper and the many suggestions (including  a few key-references on classical works) that has enabled to improve the paper.

\bibliographystyle{mn2e}


\appendix
\onecolumn

\section{The limit of small ellipticities}
\label{sec:details}

For $\epsilon_1^2 \ll 1$ and $\epsilon_2^2 \ll 1$, we can expand the $A_i$'s, and subsequently the leading term $f_1 {\cal P}(\epsilon,\epsilon')$ and the $1$st-order correction $f_1 {\cal C}(\epsilon,\epsilon')$. We have
\begin{flalign}
{\cal M}(\epsilon_1) = \frac{2}{15} \epsilon_1^2 \left( 2 + \frac{2}{7}\epsilon_1^2 + \dots \right),
\end{flalign}
and
\begin{flalign}
{\cal M}(\epsilon_1) {\cal P}(\epsilon_1,\epsilon_2) &= \frac{2}{15} \left( 3 \epsilon_2^2 -5 \epsilon_1^2 + \frac{12}{7}q^4\epsilon_2^4 - 2 \epsilon_1^2\epsilon_2^2 + \dots \right).
\end{flalign}
According to (\ref{eq:feprimab}), (\ref{eq:omega2}) and (\ref{eq:omega2correction}), we find at point B$_2$
\begin{flalign}
{\cal M}(\epsilon_2) \left. f_1   \left[ {\cal P}(\epsilon_2,\epsilon_1') + {\cal C}(\epsilon_2,\epsilon_1')\right]\right|_{{\rm B}_2} &= \left.\frac{q^3 \bareone}{\sqrt{1-{\epsilon_1'}^2}} \left[ A_0(\epsilon_1') x -A_1(\epsilon_1')\right]\right|_{{\rm B}_2}\\
\nonumber
&=  q^3 \bareone \left.\left[ \frac{A_0(\epsilon_1')}{\sqrt{1-{\epsilon_1'}^2}}  -\frac{A_1(\epsilon_1')}{\sqrt{1-{\epsilon_1'}^2}} \right]\right|_{{\rm B}_2}\\
\nonumber
&=  q^3 \bareone \left.\left[ \frac{4}{3}+ \frac{2}{15}{\epsilon_1'}^2+\frac{3}{70}{\epsilon_1'}^4 + \dots\right]\right|_{{\rm B}_2}\\
\nonumber
&=  q^3 \bareone \left[ \frac{4}{3}+ \frac{2}{15}q^2 \epsilon_1^2+\frac{3}{70}q^4\epsilon_1^4 + \dots\right],
\end{flalign}
since $\epsilon_1'|_{{\rm B}_2}=q \epsilon_1$. In a similar way, we have at point A$_2$
\begin{flalign}
{\cal M}(\epsilon_2) \left. f_1   {\cal C}(\epsilon_2,\epsilon_1') \right|_{{\rm A}_2} &=  \frac{q^3 \bareone}{(1+c)\sqrt{1+c-q^2 \epsilon_1^2}} \left. \left[ A_0(\epsilon_1') (1+c) -(1- \epsilon_2^2)A_3(\epsilon_1')\right]\right|_{{\rm A}_2},
\end{flalign}
where  $\epsilon_1'|_{{\rm A}_2}=\frac{q \epsilon_1}{\sqrt{1+c}}$ and $c$ is given by (\ref{eq:confocalc}). This latter relationship can be expanded as
\begin{flalign}
{\cal M}(\epsilon_2) \left. f_1 {\cal C}(\epsilon_2,\epsilon_1') \right|_{{\rm A}_2} &=  \left. \frac{q^3 \bareone}{(1+c)^{3/2}} \left[ \frac{A_0(\epsilon_1')}{\sqrt{1-{\epsilon_1'}^2}} (1+c) - (1 +c - q^2\epsilon_1^2)\frac{A_3(\epsilon_1')}{\sqrt{1-{\epsilon_1'}^2}}\right]\right|_{{\rm A}_2}\\
\nonumber
&=  \left. \frac{q^3 \bareone}{\sqrt{1+c}} \left[ \frac{A_0(\epsilon_1')}{\sqrt{1-{\epsilon_1'}^2}} - (1 - {\epsilon_1'}^2)\frac{A_3(\epsilon_1')}{\sqrt{1-{\epsilon_1'}^2}}\right]\right|_{{\rm A}_2}\\
\nonumber
&= \left. \frac{q^3 \bareone}{\sqrt{1+c}} \left[ \frac{4}{3} - \frac{4}{15}{\epsilon_1'}^2 - \frac{27}{70}{\epsilon_1'}^4 + \dots  + {\epsilon_1'}^2 \left( \frac{2}{3} + \frac{3}{5}{\epsilon_1'}^2 + \frac{15}{28}{\epsilon_1'}^4 + \dots  \right)\right]\right|_{{\rm A}_2}\\
\nonumber
&= \left. \frac{q^3 \bareone}{\sqrt{1+c}} \left( \frac{4}{3} +\frac{2}{5}{\epsilon_1'}^2 + \frac{3}{14}{\epsilon_1'}^4 + \dots  \right) \right|_{{\rm A}_2}\\
\nonumber
&= q^3 \bareone \left[ \frac{4}{3} \left(1-\frac{1}{2}c+\frac{3}{8}c^2 + \dots \right) +\frac{2}{5}q^2 \epsilon_1 ^2\left(1-\frac{3}{2}c
+ \dots \right) + \frac{3}{14}q^4 \epsilon_1^4\left(1
+ \dots \right) + \dots  \right].\\
\nonumber
\end{flalign}
It follows that
\begin{flalign}
{\cal M}(\epsilon_2) \left[\left. f_1 {\cal P}(\epsilon_2,\epsilon_1')\right|_{{\rm B}_2} + \left. f_1  {\cal C}(\epsilon_2,\epsilon_1')\right|^{{\rm B}_2}_{{\rm A}_2} \right]  &= \frac{2}{15} q^3 \bareone \left( 3 q^2\epsilon_1 ^2 -5 \epsilon_2 ^2 - \frac{15}{28}q^4\epsilon_1 ^4 +3 q^2 \epsilon_1 ^2\epsilon_2^2 - \frac{15}{4}\epsilon_2^4  + \dots \right)\\
\nonumber
&= \frac{2}{15} q^3 \left(1-\frac{1}{2}\epsilon_1^2-\frac{3}{8}\epsilon_1^4+\dots \right)\left( 3 q^2\epsilon_1 ^2 -5 \epsilon_2 ^2 - \frac{15}{28}q^4\epsilon_1 ^4 +3 q^2 \epsilon_1 ^2\epsilon_2^2 - \frac{15}{4}\epsilon_2^4  + \dots \right)\\
\nonumber
&= \frac{2}{15} q^3 \left[ \underbrace{3 q^2\epsilon_1 ^2 -5 \epsilon_2 ^2}_{\text{order 0}} - \frac{3}{28}q^2\epsilon_1 ^4 \left(14+5q^2\right) + \frac{1}{2}\epsilon_1 ^2\epsilon_2^2 (5+6 q^2) - \frac{15}{4}\epsilon_2^4  + \dots \right].
\end{flalign}

From these expressions, the rotations rates for the host and for the embedded spheroid are found from (\ref{eq:omega2}) and (\ref{eq:omega1v}) respectively. In these conditions, and with $\sqrt{k}=\epsilon_2/\epsilon_1$, the unique value of the mass-density jump leading to global rotation (type-C solution) is found from (\ref{eq:mdjump}), namely
\begin{flalign}
  \alphac \approx 1+ \frac{5(k-1)+ \frac{2}{7} k^2 \epsilon_1^2-2k\epsilon_1^2\left(1-\frac{6}{7}k\right)}{2\left(1+\frac{1}{7}\epsilon_1^2\right)+q^3\left[3q^2-5k-\frac{3}{2}q^2\epsilon_1^2 \left(1+\frac{5}{14}q^2\right)+ \frac{1}{15} k \epsilon_1^2(6q^2+5)-\frac{15}{4} k^2 \epsilon_1^2\right]}.
  \label{eq:slowrotc_alpha_1rstorder}
\end{flalign}

\newpage
\section{A basic F90 program}
\label{sec:f90}
\begin{verbatim}
Program nsfoe ! gfortran nsfoe.f90; ./a.out
  Implicit None
  Integer,Parameter::AP=Kind(1.00D+00)
  Real(Kind=AP),Parameter::PI=Atan(1._AP)*4
  Real(Kind=AP)::e12,e1,e1bar,e22,e2,e2bar,q,qe1,c,correction,xa,xb,fa,fb,eprima,eprimb
  Real(Kind=AP)::alpha,alphac,const1,const2,om1over2,om2over2,pif,pc,mass,vol
  ! Statements
  print*,"from J.M.Hur\'e (2021), MNRAS, 'Nested Spheroidal Figures of Equilibrium. I'"
  e2bar=0.3_AP;e1bar=0.75_AP;q=0.2_AP ! configuration F
  e2bar=0.35_AP;e1bar=0.25_AP;q=0.9_AP ! configuration E
  e2bar=0.75_AP;e1bar=0.2_AP;q=0.8_AP ! configuration D
  e2bar=0.90_AP;e1bar=0.95_AP;q=0.45_AP/e1bar;alpha=6.355758519789902_AP ! configuration A
  e22=1._AP-e2bar**2;e2=sqrt(e22);e12=1._AP-e1bar**2;e1=sqrt(e12);qe1=q*e1
  c=qe1**2-e22; print*,"Confocal parameter c",c
  xb=1._AP;eprimb=qe1/Sqrt(xb);fb=q**3*e1bar/xb/sqrt(xb-qe1**2)
  correction=fb*(cteA0(eprimb)-(1._AP-e22)*cteA3(eprimb))
  xa=1._AP+c;eprima=qe1/Sqrt(xa);fa=q**3*e1bar/xa/sqrt(xa-qe1**2)
  correction=correction-fa*(cteA0(eprima)*xa-(1._AP-e22)*cteA3(eprima))
  print*,"1rst-order correction M(e2).f.C|",correction
  pif=cteA0(e2)+(alpha-1._AP)*cteA0(e1)*q**2-(cteA3(e2)+(alpha-1._AP)*cteA3(e1))*e1bar**2*q**2
  pif=pif-(cteA0(e2)+(alpha-1._AP)*fa*cteA0(eprima)*xa)&
      &+(cteA3(e2)+(alpha-1._AP)*fa*cteA3(eprima))*e2bar**2
  pc=pif+alpha*(cteA3(e2)+(alpha-1._AP)*cteA3(e1))*e1bar**2*q**2
  print*,"Interface pressure p*(E1)",pif;print*,"Central pressure",pc 
  const1=pc/alpha-(cteA0(e2)+(alpha-1._AP)*cteA0(e1)*q**2)
  const2=-(cteA0(e2)+(alpha-1._AP)*fa*cteA0(eprima)*xa)&
      &+(cteA3(e2)+(alpha-1._AP)*fa*cteA3(eprima))*e2bar**2
  print*,"Const ",const1;print*,"Const'",const2
  alphac=1._AP+(cteM(e2)+cteM(e1)*Pos(e1,e2))/(cteM(e1)+cteM(e2)*fb*Pos(e2,eprimb)+correction)
  om1over2=cteM(e1)*(alphac-1._AP-Pos(e1,e2))
  print*,"TYPE-C SOLUTION";print*," Mass density jump",alphac;print*," Rotation rate W^2",om1over2
  print*,"TYPE-V SOLUTION";print*,"Alpha",alpha
  om2over2=cteM(e2)-(alpha-1._AP)*(fb*cteA0(eprimb)-fa*cteA0(eprima)*(1._AP+c)&
       &+fa*cteA3(eprima)*(1._AP-e22)-fb*cteA1(eprimb))
  om2over2=cteM(e2)*(1._AP-(alpha-1._AP)*fb*Pos(e2,eprimb)-(alpha-1._AP)*correction/cteM(e2))
  om1over2=(om2over2+(alpha-1._AP)*(cteA1(e2)+(alpha-1._AP)*cteA1(e1)&
       &-(cteA3(e2)+(alpha-1._AP)*cteA3(e1))*(1._AP-e12)))/alpha
  print*," Rotation rate W^2 (host)",om2over2;print*," Rotation rate W^2 (embedded ell.)",om1over2
  vol=PI*e2bar*4/3;print*,"Volume",vol;mass=PI*(e2bar+(alpha-1)*q**3*e1bar)*4/3
  print*,"Total mass",mass;print*,"Fractional mass (embedded ell.)",PI*alpha*q**3*e1bar*4/3/mass
Contains
  Function cteA0(e)
    Implicit none;Real(Kind=AP)::e,cteA0
    cteA0=2._AP;If (e>0._AP) cteA0=Sqrt(1._AP-e**2)/e*Asin(e)*2
  End Function cteA0
  Function cteA1(e)
    Implicit none;Real(Kind=AP)::e,cteA1
    cteA1=2._AP/3;If (e>0._AP) cteA1=Sqrt(1._AP-e**2)/e**2*(Asin(e)/e-Sqrt(1._AP-e**2))
  End Function cteA1
  Function cteA3(e)
    Implicit none;Real(Kind=AP)::e,cteA3
    cteA3=2._AP/3;If (e>0._AP) cteA3=Sqrt(1._AP-e**2)/e**2*(1._AP/Sqrt(1._AP-e**2)-Asin(e)/e)*2
  End Function cteA3
  Function cteM(e)
    Implicit none;Real(Kind=AP)::e,cteM
    cteM=0._AP;If (e*(1._AP-e)>0._AP) cteM=cteA1(e)-(1._AP-e**2)*cteA3(e)
  End Function cteM
  Function Pos(x,y)
    Implicit none;Real(Kind=AP)::x,y,Pos
    Pos=-1._AP;If (x*(1._AP-x)>0._AP) Pos=(cteA3(y)*(1._AP-x**2)-cteA1(y))/cteM(x)
  End Function Pos
End Program nsfoe
\end{verbatim}
\twocolumn

\end{document}